\documentclass[journal,comsoc]{IEEEtran}
\usepackage[T1]{fontenc}

\usepackage{url}
\usepackage{color}
\usepackage[square, comma, sort&compress, numbers]{natbib}
\usepackage{makecell}
\ifCLASSINFOpdf
   \usepackage[pdftex]{graphicx}
\else
\fi
\usepackage{amsmath}
\usepackage[cmintegrals]{newtxmath}
\usepackage{algorithmic}

\usepackage{booktabs}
\usepackage{multirow}

\ifCLASSOPTIONcompsoc
 \usepackage[caption=false,font=normalsize,labelfont=sf,textfont=sf]{subfig}
\else
 \usepackage[caption=false,font=footnotesize]{subfig}
\fi
\begin{document}
%
\title{Sub-sampled Cross-component Prediction for Emerging Video Coding Standards}
%
%
%

\author{Junru~Li, Meng~Wang, 
        Li~Zhang, Shiqi~Wang,~\IEEEmembership{Member, IEEE,} Kai~Zhang\\ 
        Shanshe~Wang,~\IEEEmembership{Member, IEEE}
		Siwei~Ma,~\IEEEmembership{Senior Member, IEEE}
        and  Wen~Gao,~\IEEEmembership{Fellow, IEEE}

\thanks{ J. Li, S. Wang, S. Ma and W. Gao are with the School of Electronics Engineering and Computer Science, Institute of Digital Media, Peking University, Beijing, 100871, China, (e-mail: junru.li@pku.edu.cn; sswang@pku.edu.cn; swma@pku.edu.cn; wgao@pku.edu.cn). (\textit{Corresponding author: Prof. Siwei Ma})

M. Wang and S. Wang are with Department of Computer Science, City University of Hong Kong, Hong Kong, China, (e-mail: mwang98-c@my.cityu.edu.hk; shiqwang@cityu.edu.hk).

L. Zhang and K. Zhang are with the Bytedance Inc., San Diego CA. USA, (e-mail: lizhang.idm@bytedance.com; zhangkai.video@bytedance.com).
}
        

}

\maketitle

\begin{abstract}
Cross-component linear model (CCLM) prediction has been repeatedly proven to be effective in reducing the inter-channel redundancies in video compression. Essentially speaking, the linear model is identically trained by employing accessible luma and chroma reference samples at both encoder and decoder, elevating the level of operational complexity due to the least square regression or max-min based model parameter derivation. In this paper, we investigate the capability of the linear model in the context of sub-sampled based cross-component correlation mining, as a means of significantly releasing the operation burden and facilitating the hardware and software design for both encoder and decoder. In particular, the sub-sampling ratios and positions are elaborately designed by exploiting the spatial correlation and the inter-channel correlation. Extensive experiments verify that the proposed method is characterized by its simplicity in operation and robustness in terms of rate-distortion performance, leading to the adoption by Versatile Video Coding (VVC) standard and the third generation of Audio Video Coding Standard (AVS3). 
\end{abstract}

\begin{IEEEkeywords}
Cross-component linear model, VVC, AVS3, cross-component prediction, video coding.
\end{IEEEkeywords}

%
\IEEEpeerreviewmaketitle

\section{Introduction}
\IEEEPARstart{T}{here} has been a tremendous growth of interest in developing high-efficiency video coding technologies, the key of which lies in redundancy removal such that the video data can be compactly represented with justifiable quality degradation. The typical YCbCr color space~\cite{bovik2010handbook}, which reveals significant beneficial characteristics in the energy concentration, has become the most prevalent color space in video compression, transmission and display. Within YCbCr, strong redundancies exist among different color channels, motivating numerous explorations towards better representation capability by modelling the relationship between different color components.

Essentially speaking, luma channel is characterized with the main structural information, and by contrast chroma components are relatively more homogeneous. As such, the luma component is accompanied with finer prediction strategies, and the cross-component linear prediction can be further performed in an effort to infer chroma from luma. More specifically, chroma coding block (CB) $P_c^{'}$ can be predicted with the corresponding luma reconstructed CB $R_l$ through a linear model as follows,
\begin{align}
    P_c^{'} = \alpha\cdot R_l + \beta,
\end{align}
where $\alpha$ and $\beta$ are linear model parameters. By performing such cross-component prediction, the inter-channel redundancies can be effectively removed, leading to the improvement of the coding performance. To obtain the linear model parameters, instead of explicitly signaling, $\alpha$ and $\beta$ are typically obtained by the derivation with the neighboring luma and chroma reference sample pairs at both encoder and decoder through the least square regression (LSR). 

Such straightforward modelling between luma and chroma based on reconstructed samples may not be able to fully explain their complicated correlation, motivating a series of efforts devoted to improving the inference accuracy~\cite{zhangkaiTIP, zhangxingyuTIP, ZhangTao, Liyue}. In particular, they are developed from the perspective of expanding the potential reference regions, sophisticated mapping with piecewice linear and hybrid neural network. These strategies usually bring in performance improvement at the expense of highly increased complexity. Another vein is developing low complexity models, relying on the selection of the training samples~\cite{M0263, M0274, M0356, M0401,maxmin}. In particular, the Max-Min~\cite{maxmin} based cross-component prediction was developed based on the principle that sample pairs with the maximum and minimum luma intensities are regarded to be representative in deriving the linear color mapping relationship. However, to meet the real-world application scenarios, these strategies still suffer from the non-negligible computational burdens which are doubtless unfriendly to the hardware and software implementations.

In this paper, we propose the sub-sampled cross-component prediction the principle of which is reliance on the assumption that the elaborate down-sampling could well preserve the inter-channel relationship. As such, to alleviate the issue of high complexity in cross-component inference, instead of employing the full-set of the reference sample pairs in deriving the linear model, we adopt a sub-sampled set of the reference samples where at most four sample pairs with fixed relative positions are employed. 
{In particular, this paper presents the sub-sampling strategy in an analytical way from the perspective of spatial correlation and inter-channel correlation, which is an extension of our previous work in~\cite{DCC_CCLM}.} The proposed scheme considerably reduces the operation complexity and maintains the rate-distortion performance, which has been validated through a series of experiments.
The sub-sampled cross-component prediction has been adopted by the versatile video coding (VVC) standard~\cite{vvc4}. Moreover, based on the design philosophy of sub-sampled cross-component prediction, we further propose the Two Step Cross-component Prediction Mode (TSCPM), which has been adopted in the third generation of audio and video coding standard (AVS3)~\cite{AVS3_jiaqi} chroma intra prediction. 

\begin{figure*}[t]
    \centering
    \subfloat[]{\includegraphics[width=2.5in]{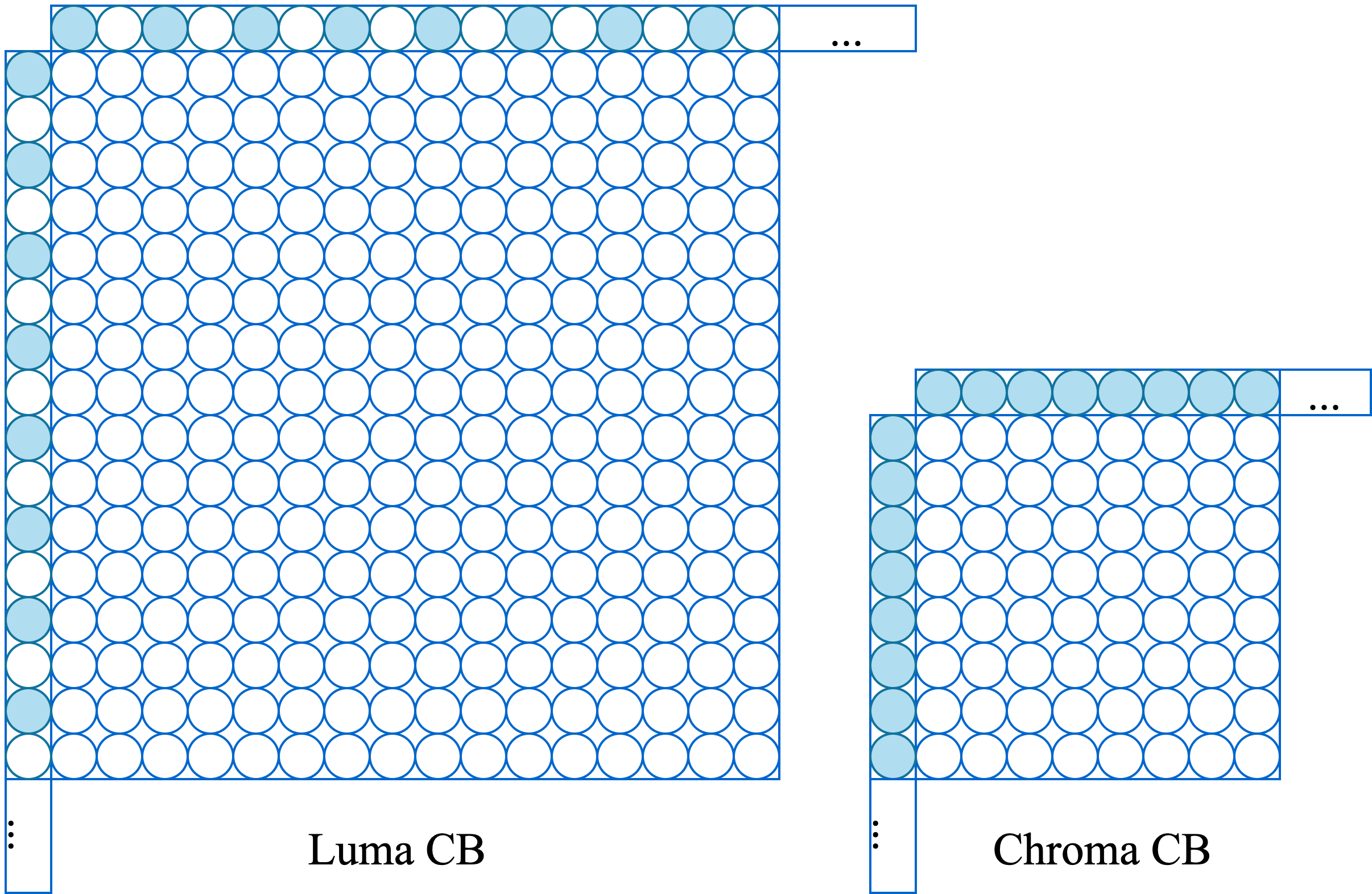}}
    \hfil
    \subfloat[]{\includegraphics[width=1.7in]{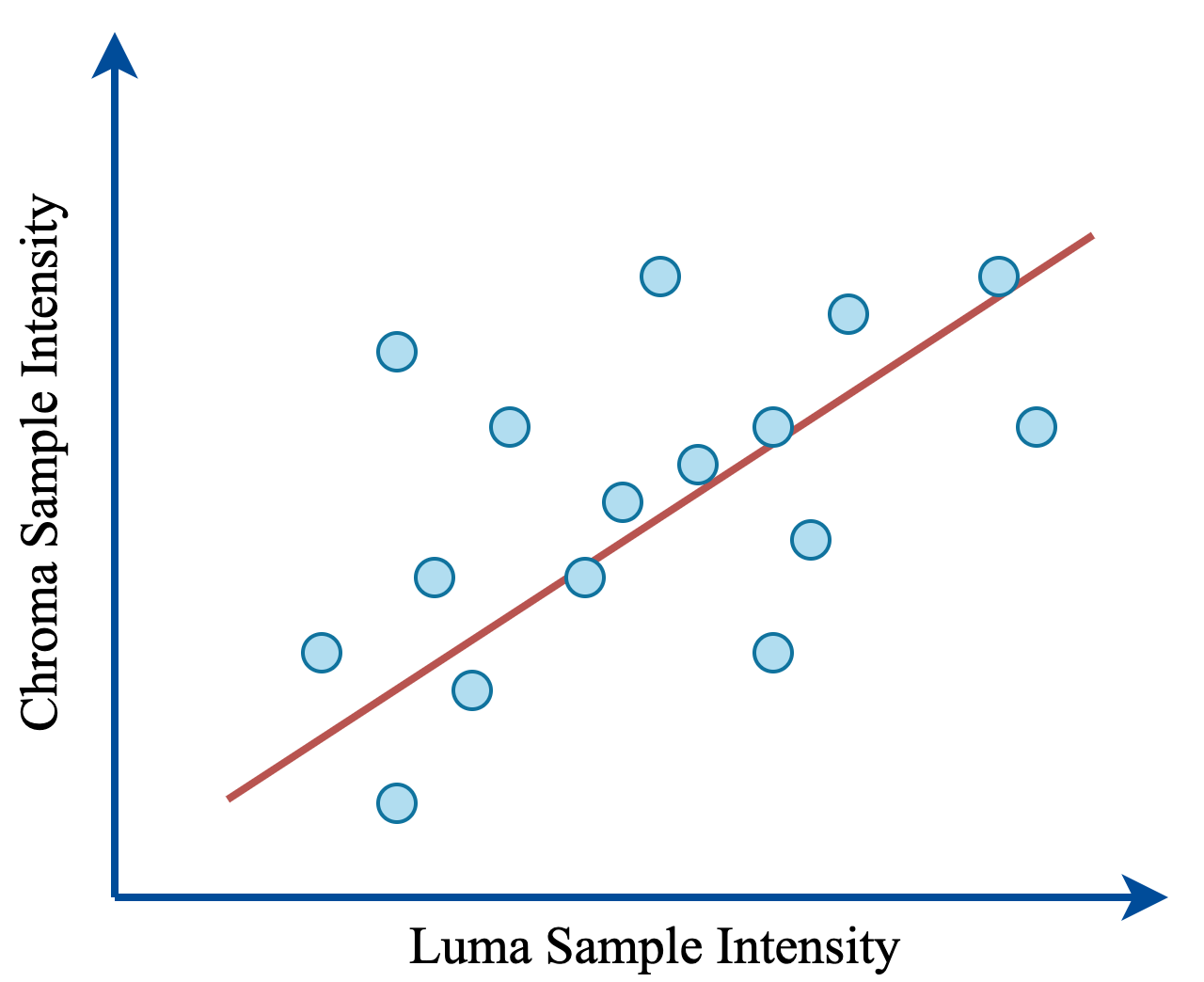}}
    \hfil
    \subfloat[]{\includegraphics[width=1.7in]{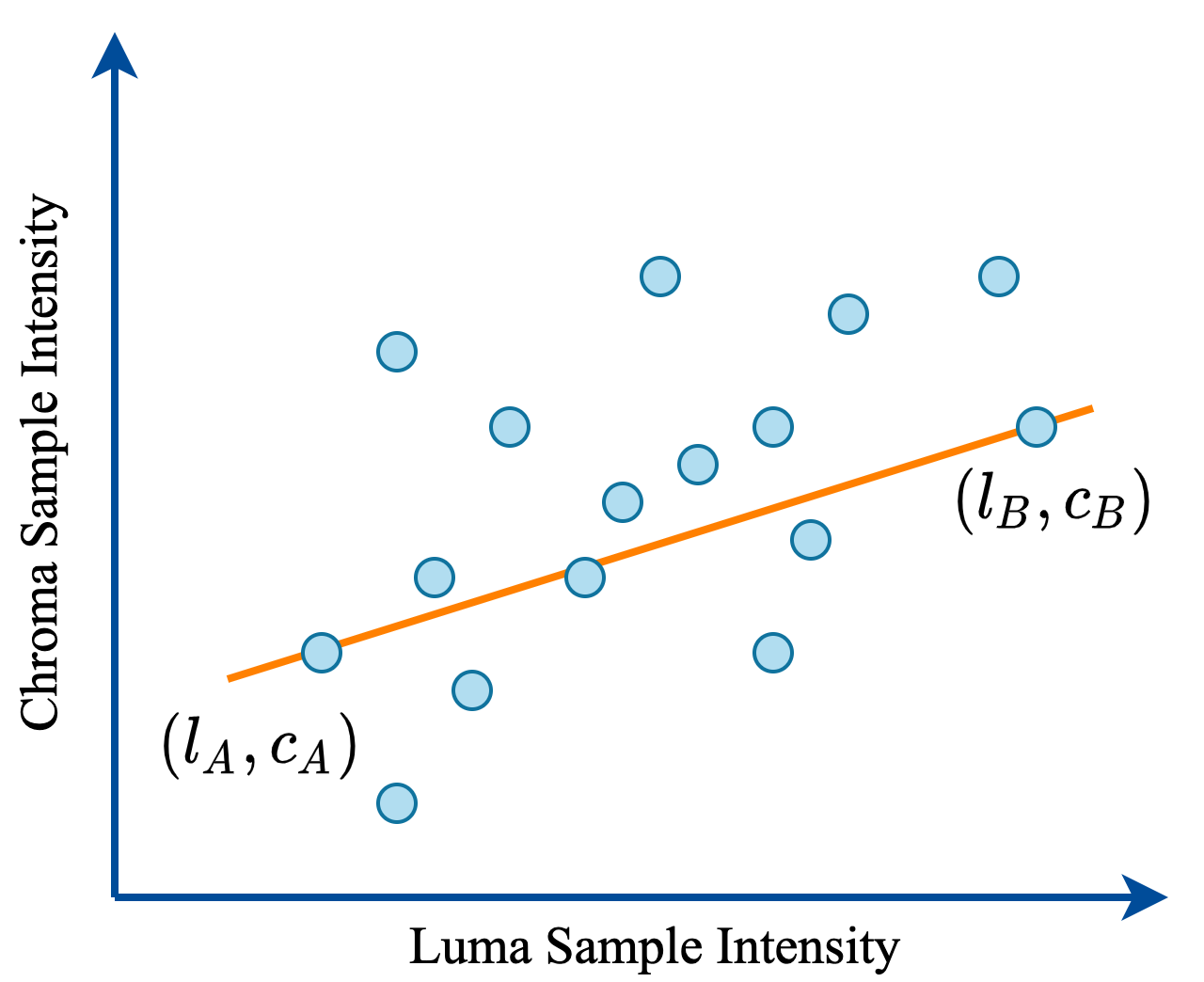}}
\caption{{Illustration of the locations of the reference sample pairs and the linear model derivations. (a) Sample positions; (b) LSR method; (c) Max-Min method~\cite{maxmin}.}}
\label{LSR_MAXMIN}
\end{figure*}
 
\section{Related Works}
The development of video coding technologies are driven by the video coding standards which typically enjoy the classical hybrid coding framework. Within this framework, numerous video coding tools have been extensively investigated. In particular, most of them share the identical principle that better removal of the redundancies within video data could lead to higher compression performance. In particular, in the upcoming VVC~\cite{vvc4} and the AVS3~\cite{AVS3_jiaqi} launched in 2018, a series of advanced techniques have been developed, in an effort to better exploit the redundancies within visual signals. More specifically, to determine the scale space in prediction, flexible coding unit partitioning structures, such as quad-tree nested binary tree and ternary tree~\cite{MTT}, as well as extended quad-tree~\cite{EQT_TIP} are employed in VVC and AVS3, with the goal of promoting local adaptability. Moreover, enhanced intra and inter prediction technologies~\cite{AMC, ISP_VVC, WideAngular, AMVR, HMVP_ICME, HMVP_DCC, FIMC_TIP} have been investigated to improve the prediction efficiency with the consideration of various video characteristics. In this manner, the signal-level redundancy~\cite{AMC, ISP_VVC, WideAngular, AMVR}, as well as the mode-level redundancy~\cite{HMVP_ICME, HMVP_DCC,FIMC_TIP} can be elegantly removed. In terms of transform, VVC and AVS3 support multiple transform cores such as DST-VII and DCT-VIII for better residual energy compaction~\cite{Transform_Zhaoxin}. 

In addition to the spatial and temporal redundancies, statistically there exist high correlations across different components. The cross-component prediction could be performed in the original pixel domain~\cite{CCLM2010, CCLM2010_Oct} or predicted residual domain~\cite{CCP_HEVC,rudat2019inter}, both of which have been proven to be effective in promoting the video compression efficiency. In particular, the residual domain cross-component prediction, which was adopted in HEVC Range Extensions for 4:4:4 color format~\cite{HEVCREOverview}, has revealed its benefits in maintaining high color fidelity. Regarding the prediction model, in the literature, both ordinary linear mapping~\cite{CCLM2010, zhangxingyuTIP, zhangkaiTIP, ZhangTao} and sophisticated nonlinear mapping~\cite{Liyue} have been studied. In~\cite{zhangkaiTIP}, a multi-model based cross-component linear model (CCLM) prediction was proposed to enhance the prediction efficiency, which was adopted to JEM platform~\cite{jem4}. Zhang \textit{et al.}~\cite{zhangxingyuTIP} proposed three additional CCLM modes that employ only one side of reference samples in the linear model derivation. In~\cite{ZhangTao}, the reference region was extended with the above-right and below-left reference samples when deriving the linear model, and adaptive Cr prediction scheme was studied to further reduce the inter-channel redundancies. Moreover, since a coding unit could contain a variety of colors, the diversification of the cross-component relationships can be effectively exploited by the nonlinear prediction with the hybrid neural network~\cite{Liyue}, whereas the derivation of the model parameters is highly computationally expensive. 

The emerging video compression standards VVC and AVS3 employ cross-component prediction as a new chroma intra prediction strategy, which brings significant performance improvement especially in terms of the chroma rate distortion performance. Typically, three CCLM modes are involved depending on which side of the reference samples is employed for the model derivation, including LM mode, LM-Above mode and LM-Left mode. Regarding LM mode, reference samples locating at the left neighboring column and the above neighboring row are all eligible for the derivation of the model parameters. Analogously, only the left side and the above side of the reference samples are utilized for LM-Left and LM-Above mode, respectively.

\section{Motivations} 
Typically, LSR has been employed in the conventional or the enhanced CCLM method when deriving the linear model parameters~\cite{zhangxingyuTIP, zhangkaiTIP, ZhangTao}.
Supposing there are $M$ accessible reference sample pairs, with the conventional LSR, the linear model parameters $\alpha$ and $\beta$ can be obtained as follows,
\begin{align}
    \alpha &= \frac{M\cdot \sum\limits_{m=1}^{M}L^{(m)}\cdot C^{(m)}- \sum\limits_{m=1}^M L^{(m)}\cdot\sum\limits_{m=1}^M C^{(m)}}{M\cdot\sum\limits_{m=1}^{M}(L^{(m)})^2 - (\sum\limits_{m=1}^{M}L^{(m)})^2},\nonumber\\
    \beta &= \frac{\sum\limits_{m=1}^M C^{(m)} - \alpha \cdot \sum\limits_{m=1}^M L^{(m)}}{M},
\end{align}
where $L^{(m)}$ and $C^{(m)}$ denote the luma and chroma reference samples respectively with index $m$. Regarding the 4:2:0 color format, luma down-sampling is performed to align with the dimension of chroma samples. The neighboring sample positions regarding the luma and chroma pairs, and the linear model are illustrated in Fig.~\ref{LSR_MAXMIN}(a) and Fig.~\ref{LSR_MAXMIN}(b). We analyze the operation complexity regarding the LSR in Table~\ref{operations_LSR_MaxMin}. In particular, $2M+4$ multiplications and $7M+3$ additions are involved in calculating $\alpha$ and $\beta$ with LSR. Meanwhile, down-samplings should be applied to the $M$ neighboring luma samples. Moreover, two divisions can be replaced by shifting with assistance of a look-up-table. As such, considerable number of multiplication, addition and down-sampling operations are involved 
in deriving the model parameters through LSR, which is extremely unfriendly to the hardware implementation. 

It is generally acknowledged that multiplication takes account of larger computational interval than addition, shifting, and comparison. To reduce the quantities of multiplication, the Max-Min method~\cite{maxmin} has been adopted by VVC~\cite{vtm3}, in an effort to approximate the linear relationship. With Max-Min method, only two sample pairs with the maximum and minimum luma intensities, as illustrated in Fig.~\ref{LSR_MAXMIN}(c), are involved in the calculation of $\alpha$ and $\beta$,
\begin{align}
    \alpha &= \frac{c_B - c_A}{l_B - l_A}, \nonumber \\
    \beta  &= c_A - \alpha \cdot l_A,
    \label{max_min}
\end{align}
where $l_A$ and $l_B$ denote the minimum and the maximum luma intensities, respectively. Moreover, $c_A$ and $c_B$ are corresponding chroma sample intensities. In this manner, the number of multiplications and additions can be largely reduced, as illustrated in Table~\ref{operations_LSR_MaxMin}. However, the complicated multiplications are migrated to the searching of the maximum and the minimum luma samples, which involves $2M$ comparisons in total. In 
this scenario, luma down-sampling is required to be performed on the $M$ neighboring samples. Therefore, it is unfortunate that the Max-Min method still requires excessive computational resources. 

In real-world applications, it is imperative to develop
a new CCLM model that enjoys low complexity in model parameter derivation and high efficiency in terms of rate-distortion performance. As is common in many image processing methods, we propose a sub-sampled cross-component prediction scheme relying on the down-sampled reference sample pairs, such that the number of comparisons and down-sampling operations can be feasibly controlled. Furthermore, in order to avoid the corruption of the primitive structures 
that capture the correlations between luma and chorma components, we design the sub-sampling strategy and select the most efficient set of sample pairs in a scientifically sound way, according to the inter-channel and spatial correlations. 
\begin{table*}[t]
  \centering
  \caption{Illustration of the Operation Complexities Regarding the LSR and the Max-Min Method}
    \begin{tabular}{c|c|c|c|c|c}
    \toprule
    Operations & Multiplication & Addition & Shift & Comparison & Down-sampling \\
    \midrule
    LSR~\cite{vvc2}   & 2$M$+4  & 7$M$+3  & 2     & -     & $M$ \\
    \midrule
    Max-Min~\cite{maxmin} & 1     & 3     & 1     & 2$M$    & $M$ \\
    \bottomrule
    \end{tabular}%
  \label{operations_LSR_MaxMin}%
\end{table*}%

\section{Sub-sampled Cross-component Prediction}
In this section, we first investigate the sub-sampling ratios that are sufficient in sustaining the primitive luma and chroma correlations. Subsequently, sample positions are selected in an analytical way which are compatible to different linear model prediction modes. Finally, the proposed linear model derivation method is presented, along with the discussions regarding the sensitivity of the linear model parameters.

\begin{figure*}[!htb]
    \centering
    \subfloat[]{\includegraphics[width=1.7in]{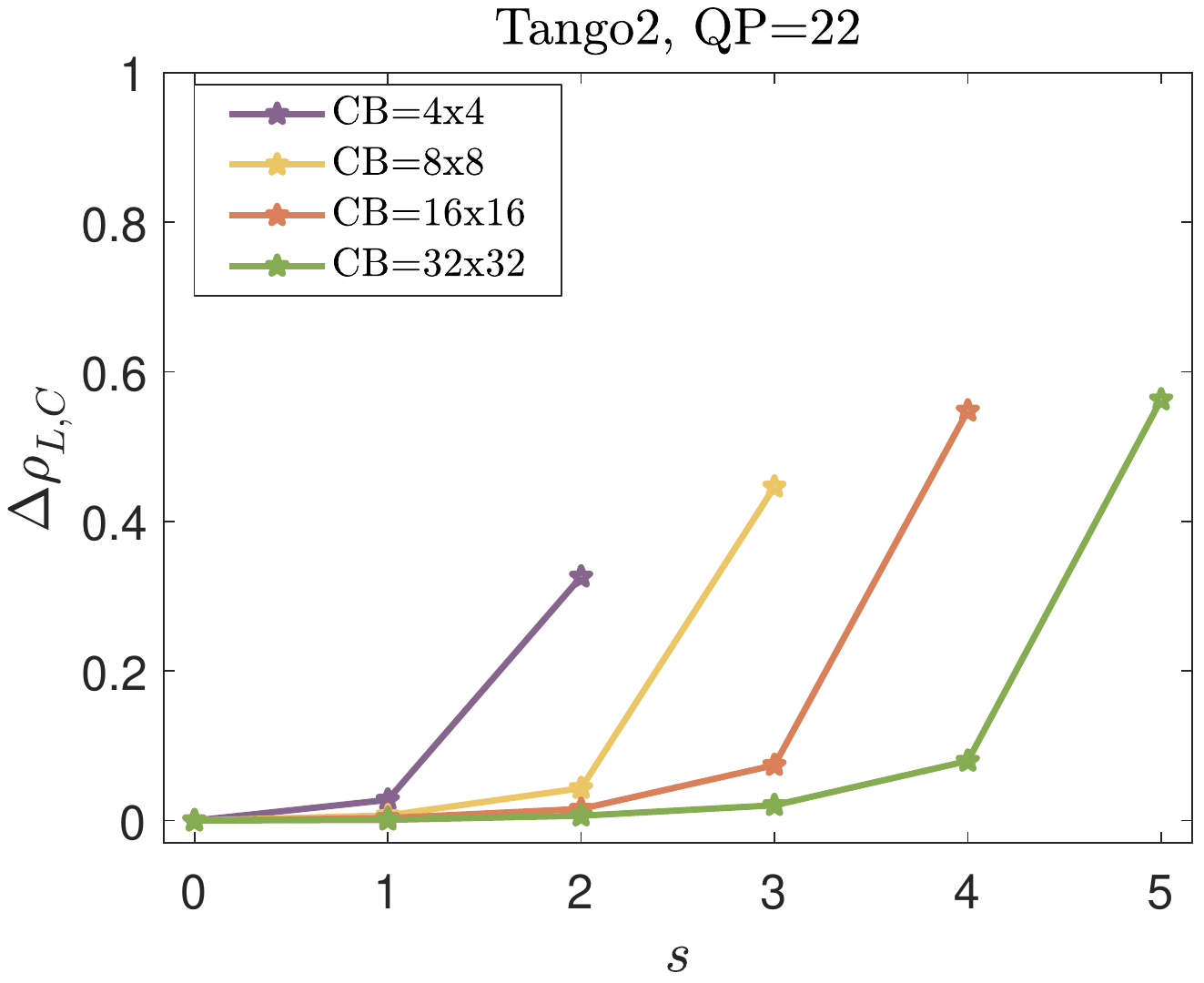}}
    \hfil
    \subfloat[]{\includegraphics[width=1.7in]{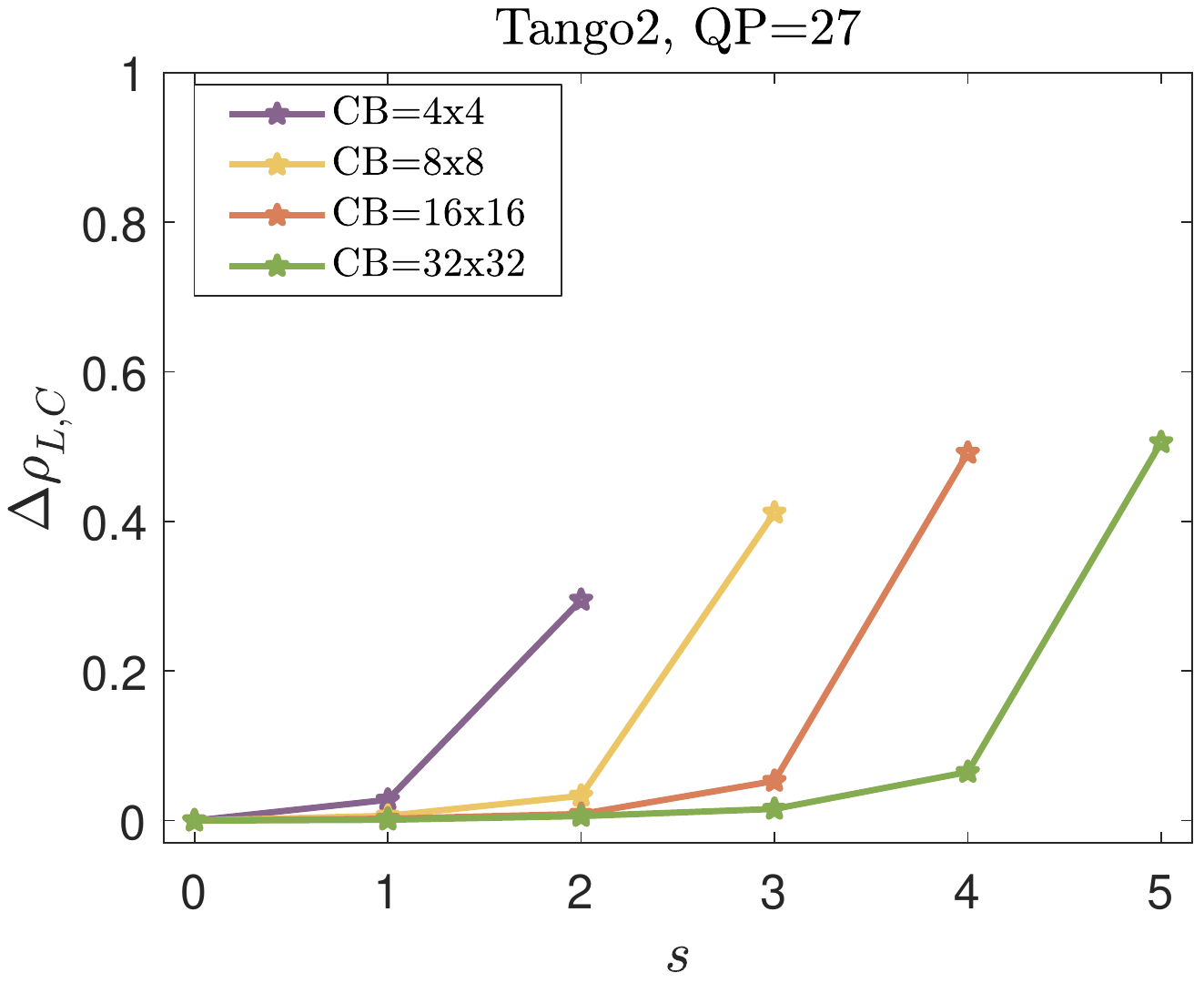}}
    \hfil
    \subfloat[]{\includegraphics[width=1.7in]{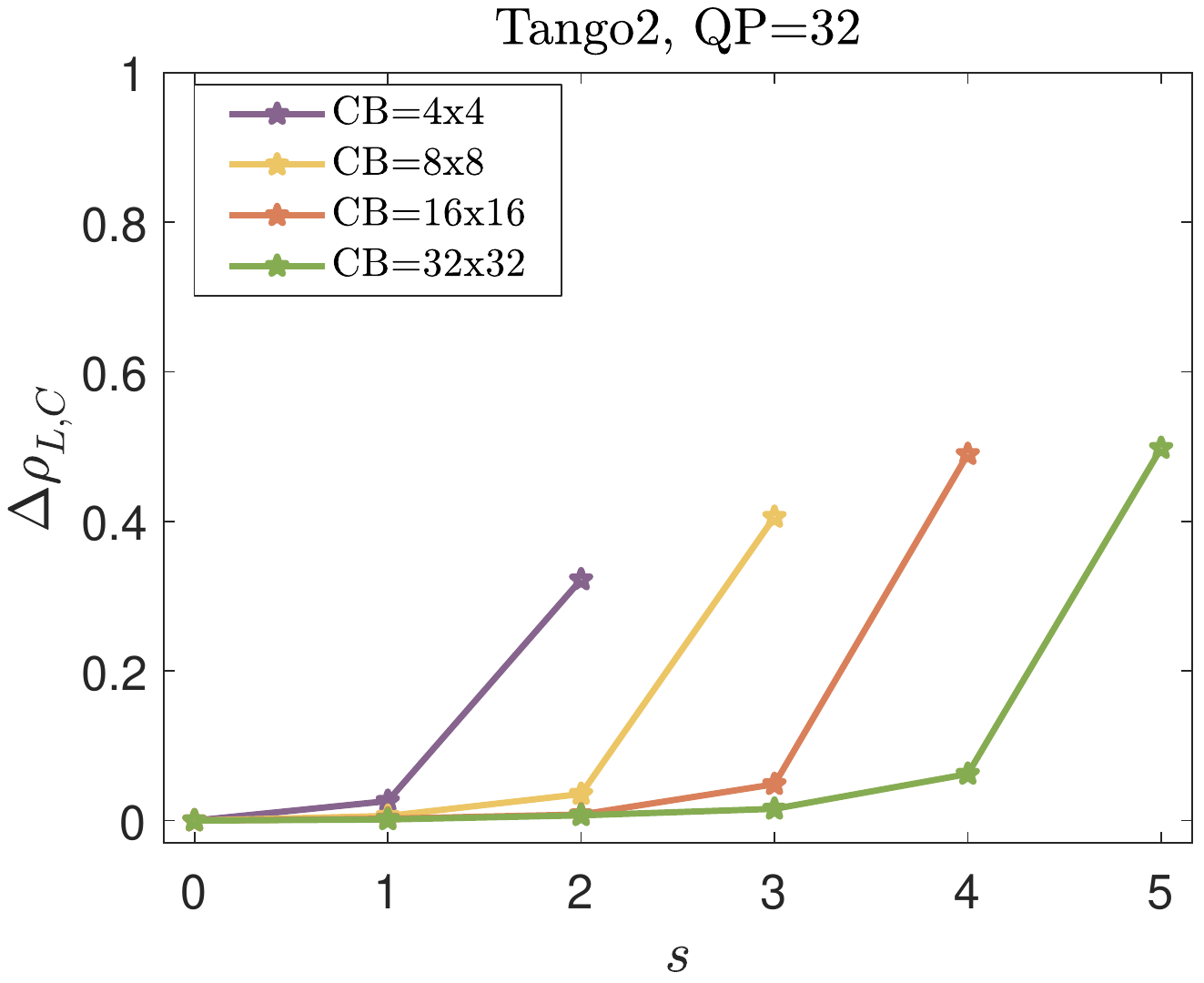}}
    \hfil
     \subfloat[]{\includegraphics[width=1.7in]{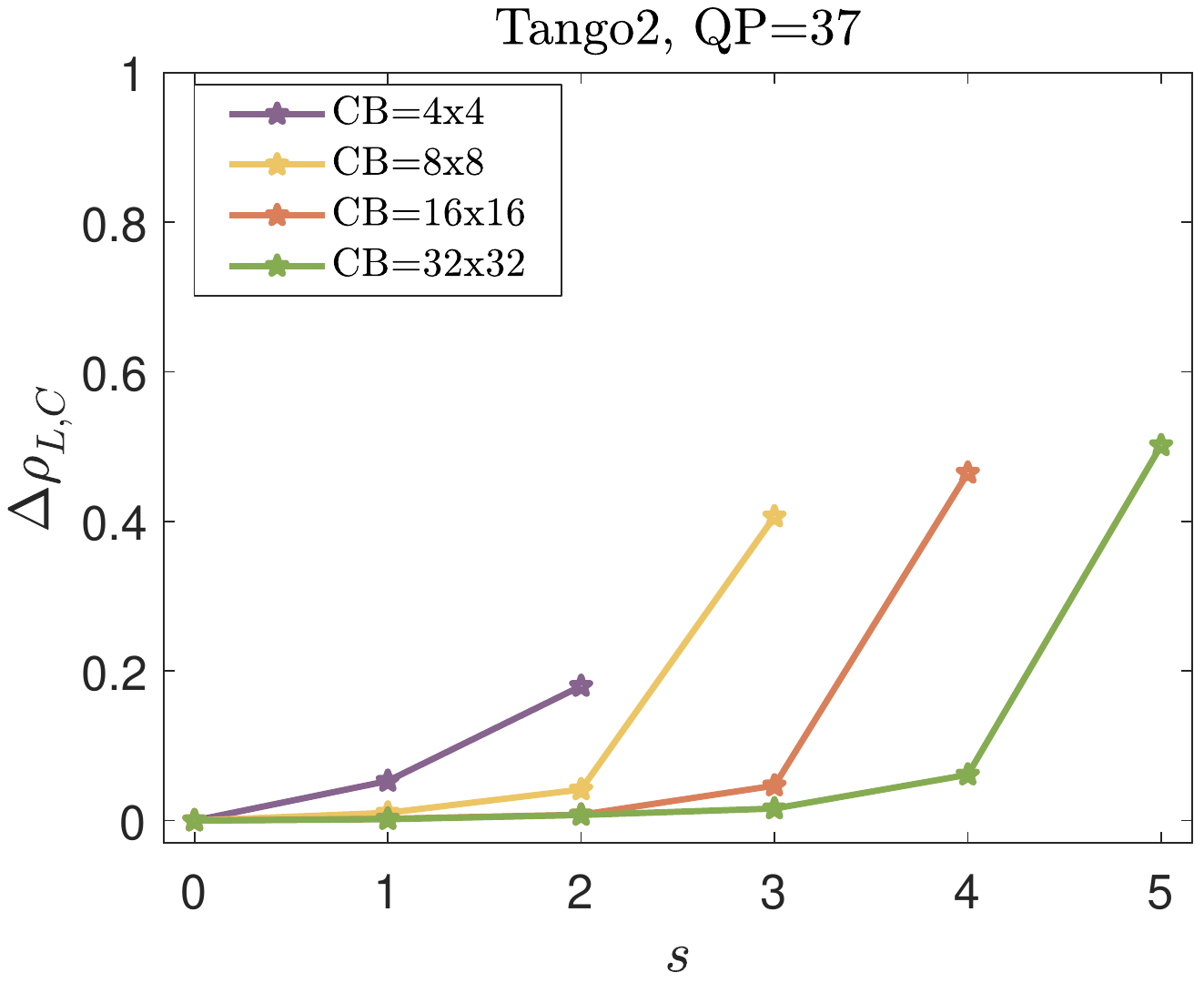}}\\
     \subfloat[]{\includegraphics[width=1.7in]{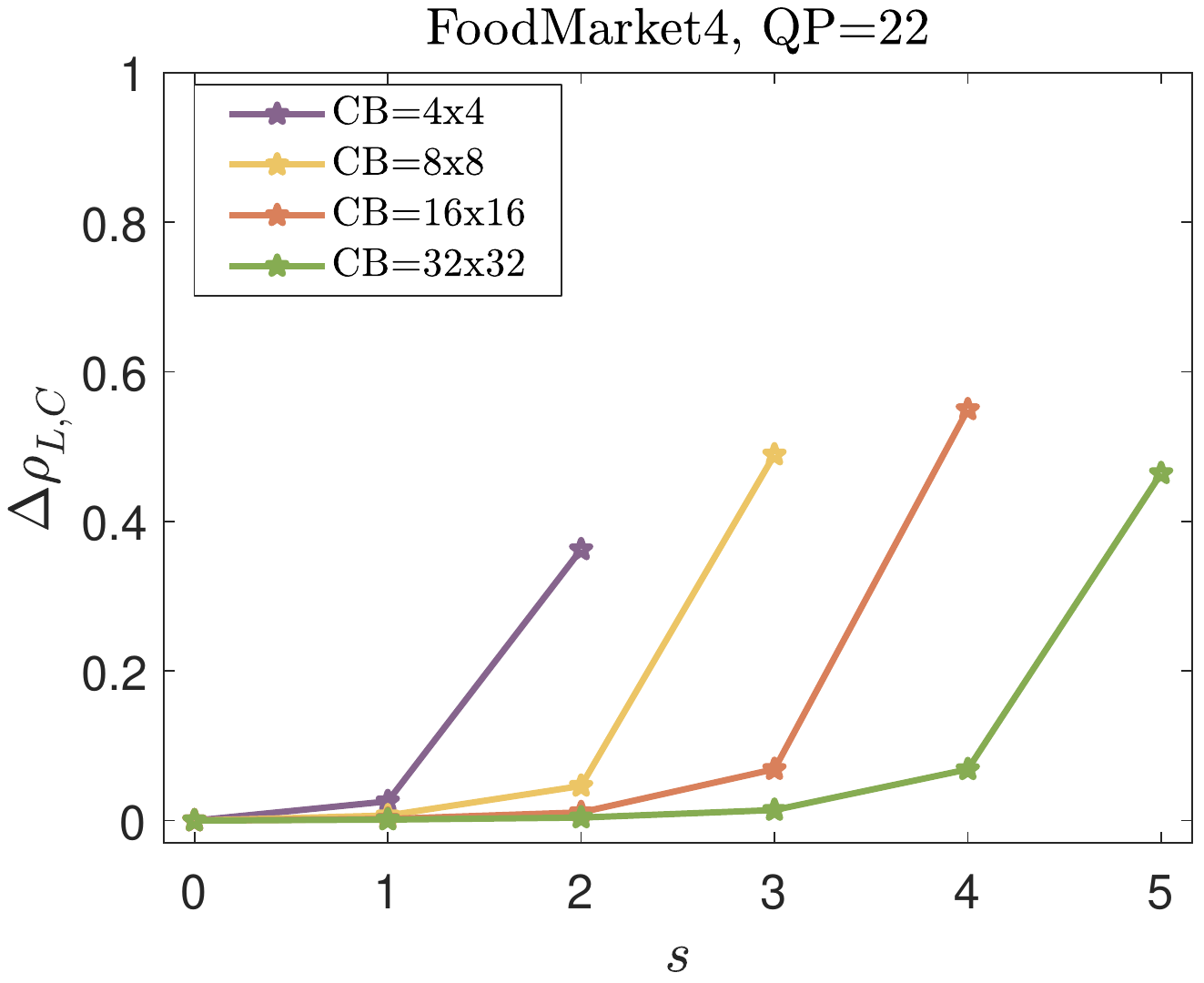}}
    \hfil
    \subfloat[]{\includegraphics[width=1.7in]{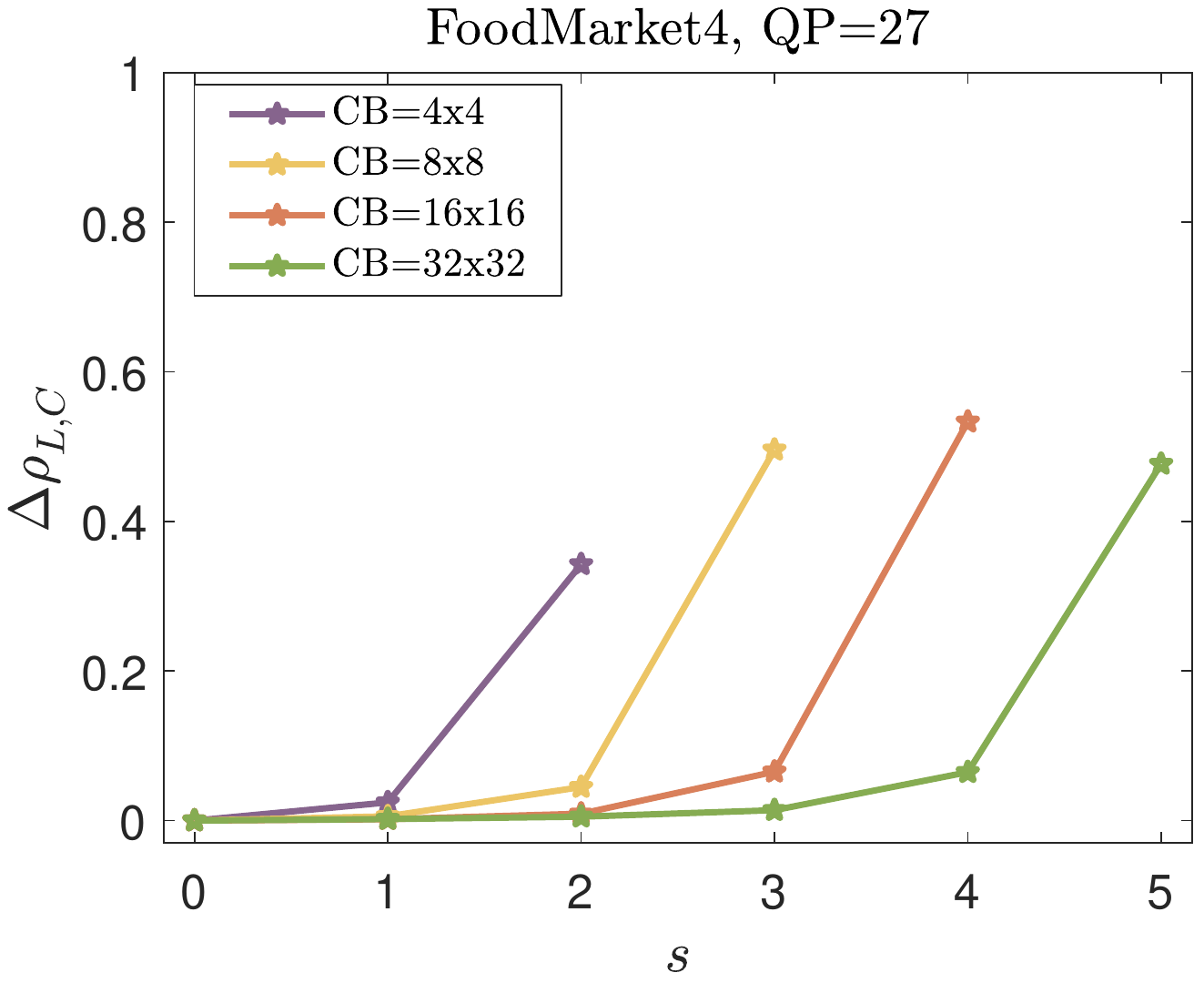}}
    \hfil
    \subfloat[]{\includegraphics[width=1.7in]{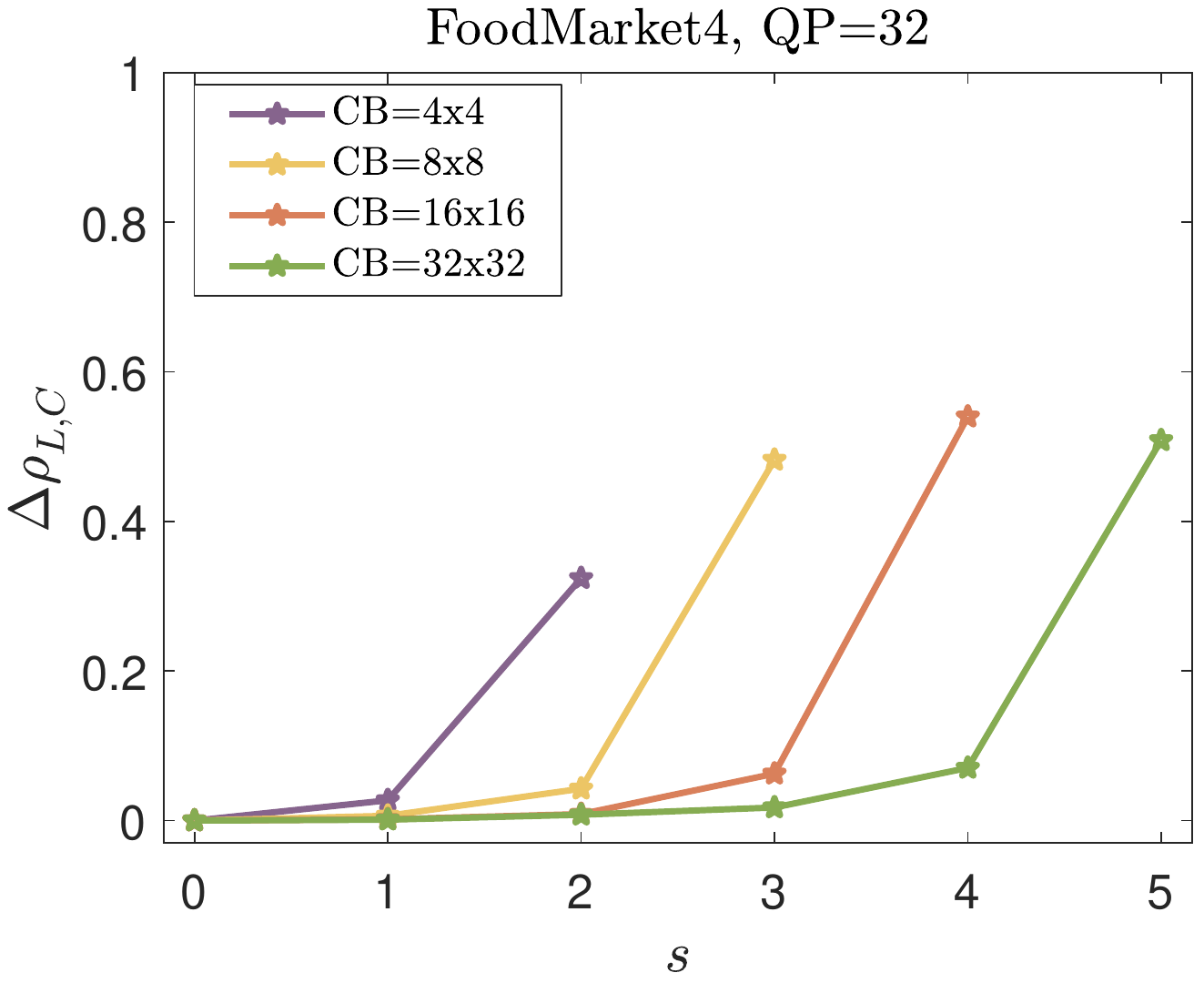}}
    \hfil
     \subfloat[]{\includegraphics[width=1.7in]{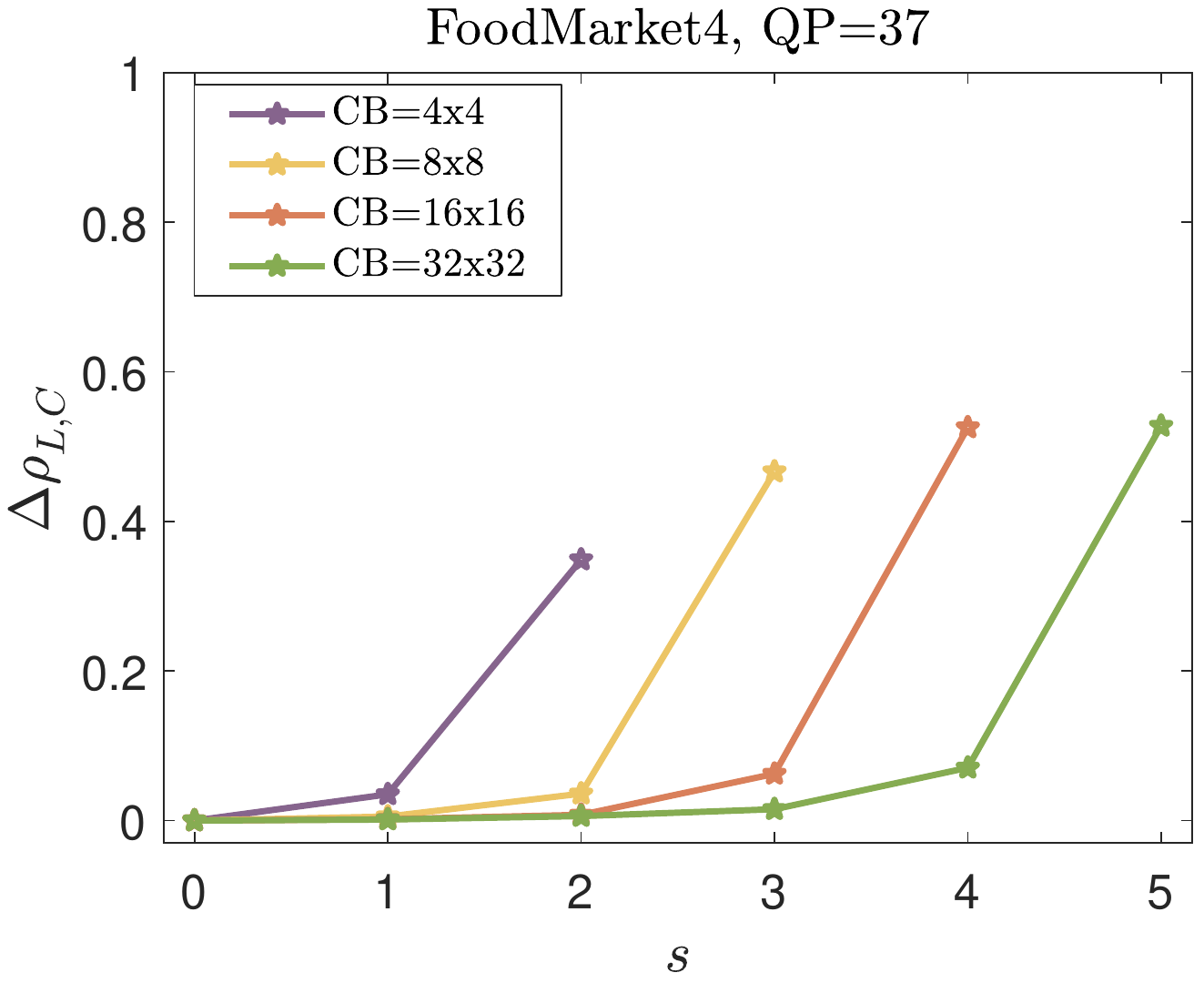}}\\
     \subfloat[]{\includegraphics[width=1.7in]{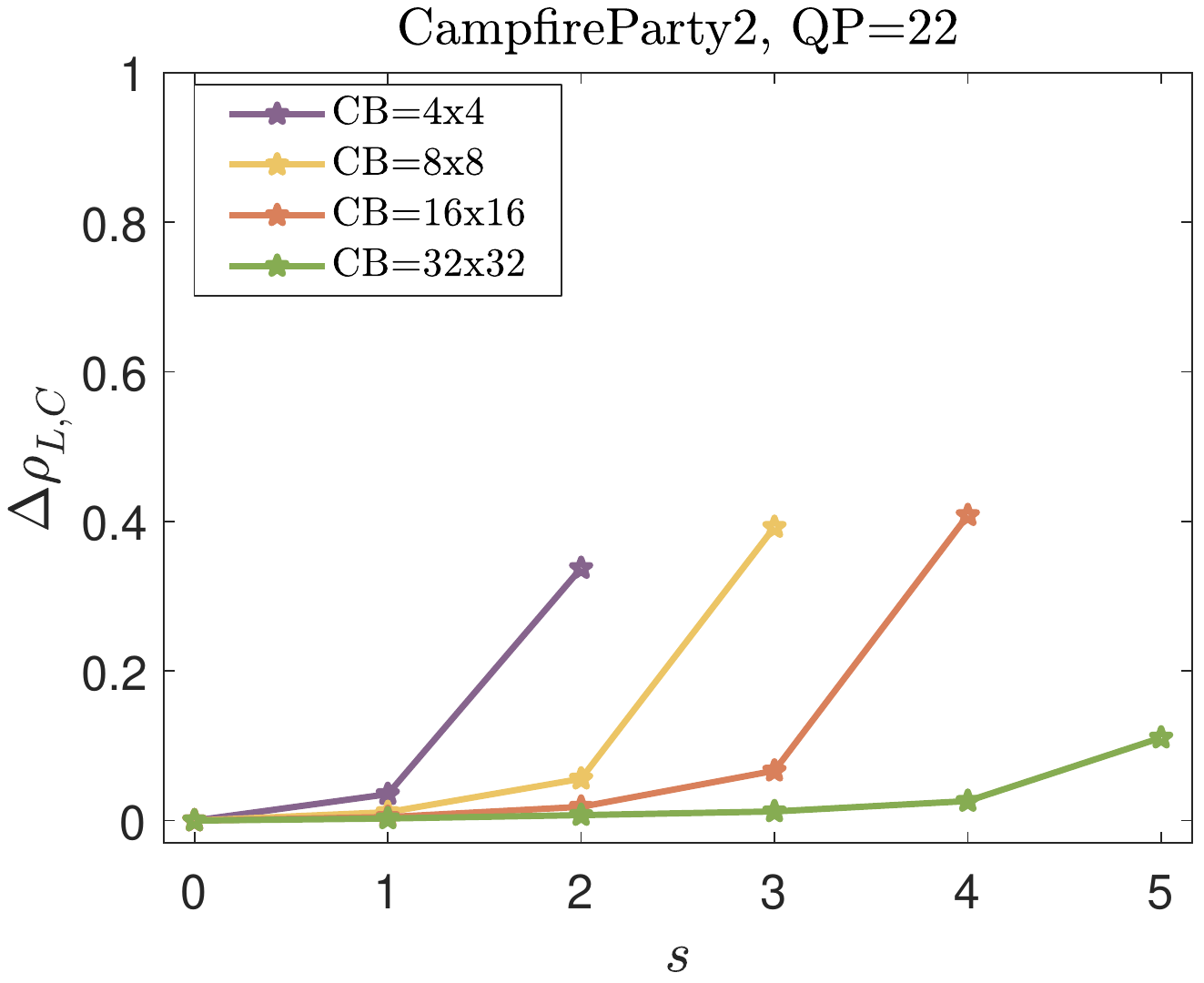}}
    \hfil
    \subfloat[]{\includegraphics[width=1.7in]{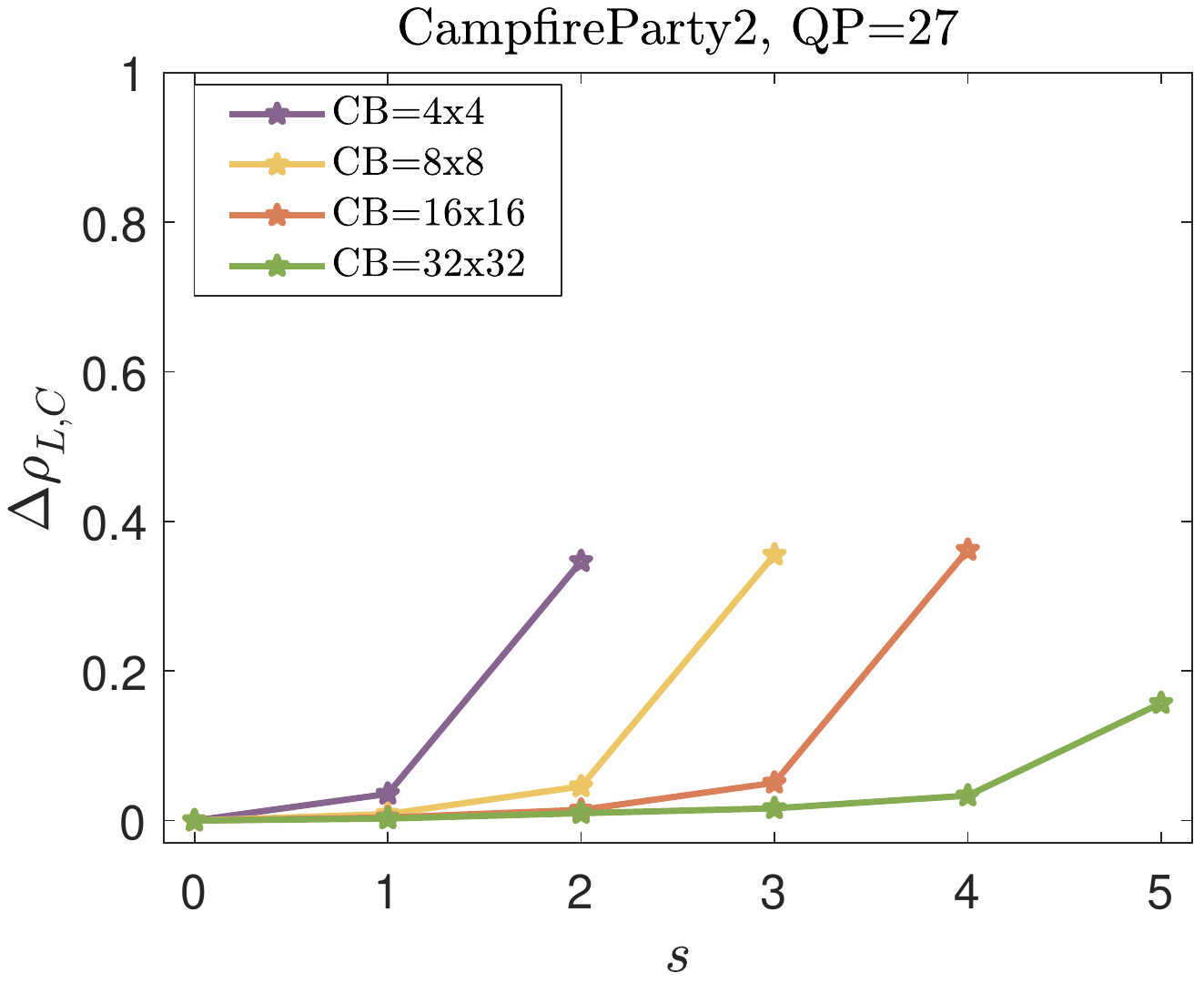}}
    \hfil
    \subfloat[]{\includegraphics[width=1.7in]{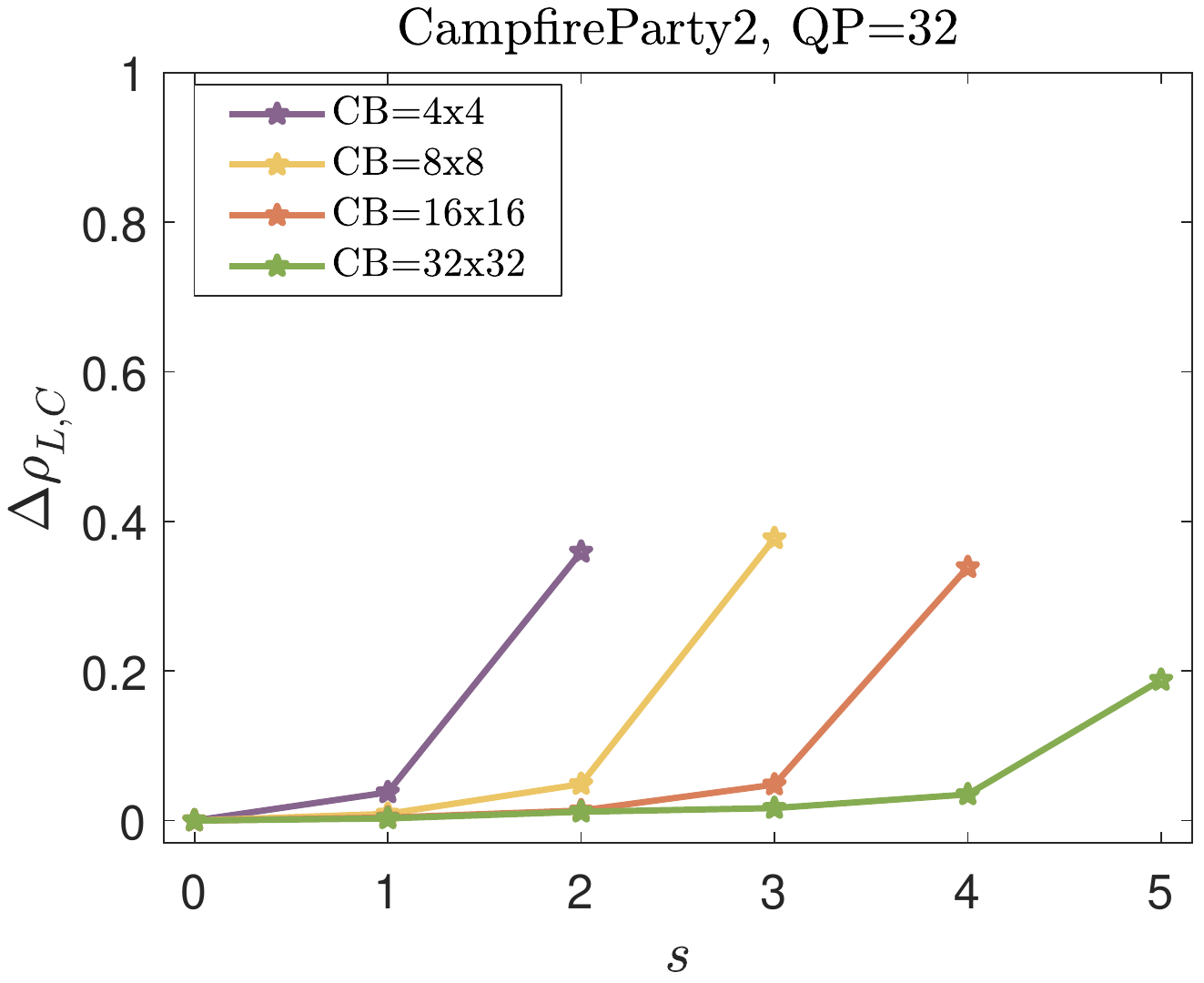}}
    \hfil
     \subfloat[]{\includegraphics[width=1.7in]{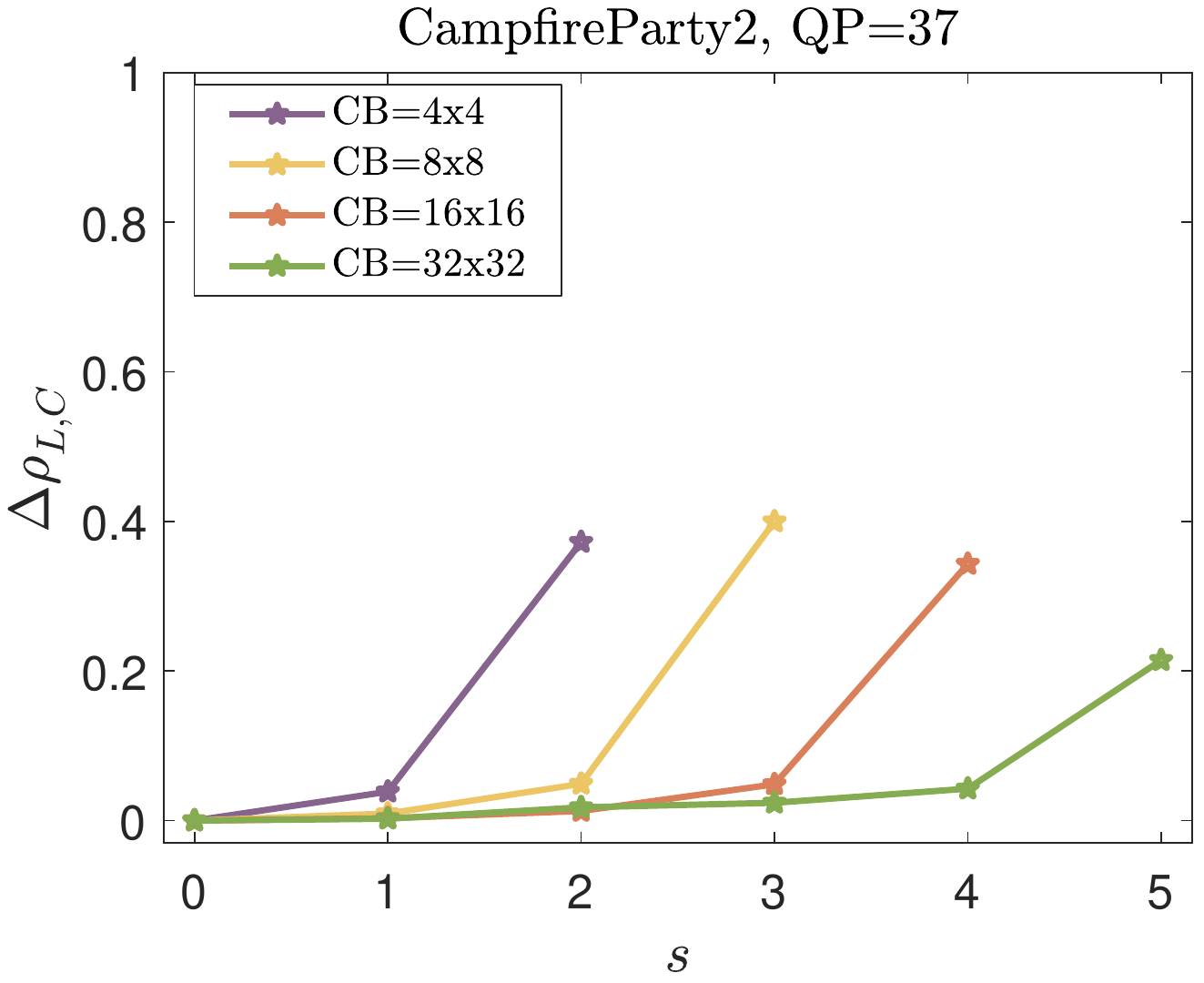}}\\
\caption{{Illustration of {$\Delta\rho_{L,C}$} with varied sub-sampling ratios $s$, {CB} sizes and QPs in sequences ``Tango2'', ``FoodMarket4'' and ``CampfireParty2''.}}
\label{deltaRhoLC}
\end{figure*}

\subsection{Sub-sampling Ratios}
Involving all accessible neighboring sample pairs in the linear model derivation will definitely increase the overhead regarding the operation and memory access, especially in terms of the hardware implementations. Therefore it is highly desirable to employ a sub-sampled set to derive the linear model parameters. However, sub-sampling might degrade the original structure, which leads to a strong correlation between luma and chroma components. More specifically, we investigate the variation of the correlation between luma component $L$ and chroma component $C$ with different sub-sampling ratios $s$. In particular, the neighboring above and left reconstructed samples are involved in analyses, which assembles the actual encoding procedure. The correlation between luma and chroma components with sub-sampling ratio $s$ can be formulated with the covariance between $L$ and $C$ as follows,
\begin{align}
    \rho_{L,C}^s=\frac{cov(L, C)}{\sigma_L \cdot \sigma_C},
\end{align}
where $\sigma_L$ and $\sigma_C$ denote the standard deviation of $L$ and $C$, respectively. $\Delta\rho_{L,C}$ is used to represent the difference of the luma and chroma correlation with respect to different sub-sampling ratios $s$ as follows,
\begin{align}
    \Delta\rho_{L,C} = \left(\rho_{L,C}^s - \rho_{L,C}^0\right)^2.
\end{align}
Assuming both the left side and the above side neighboring samples are available, the number of accessible sample pairs with different subsampling ratios $s$ is illustrated in Table~\ref{subSamplingTimes}, where only the exact above and exact left reference samples are considered. To be more specific, for a typical 32 $\times$ 32 chroma CB, there are 64 luma and chroma reference sample pairs when $s$ equals to zero. When $s$ is increased to 5, only two sample pairs are left. The $\Delta\rho_{L,C}$ with regard to different $s$ on different CB sizes is illustrated in Fig.~\ref{deltaRhoLC}. The quantization parameter (QP) varies from 22 to 37. We can observe that $\Delta\rho_{L,C}$ increases with the $s$, and the trend is independent with the QPs. Moreover, the influences at the second to the last point from the those lines suggest that four sample pairs are sufficient to maintain the existing inter-channel correlations. Therefore, in our method, at most four sample pairs are employed in the cross-component prediction.
\begin{table}[t]
  \centering
  \caption{Illustration of the Number of Luma and Chroma Sample Pairs with Different Sub-sampling Ratios $s$ in Varied Chroma CB Sizes}
    \begin{tabular}{c|cccccc}
    \toprule
    \multirow{2}[4]{*}{Chroma CB Sizes} & \multicolumn{6}{c}{$s$} \\
\cmidrule{2-7}          & 0     & 1     & 2     & 3     & 4     & 5 \\
    \midrule
    4$\times$4   & 8     & \textbf{4}     & 2     & -     & -     & - \\
    8$\times$8   & 16    & 8     & \textbf{4 }    & 2     & -     & - \\
    16$\times$16 & 32    & 16    & 8     & \textbf{4}     & 2     & - \\
    32$\times$32 & 64    & 32    & 16    & 8     & \textbf{4}     & 2 \\
    \bottomrule
    \end{tabular}%
  \label{subSamplingTimes}%
\end{table}%
\begin{figure}[t]
    \centering
    \subfloat[]{\includegraphics[width=1.7in]{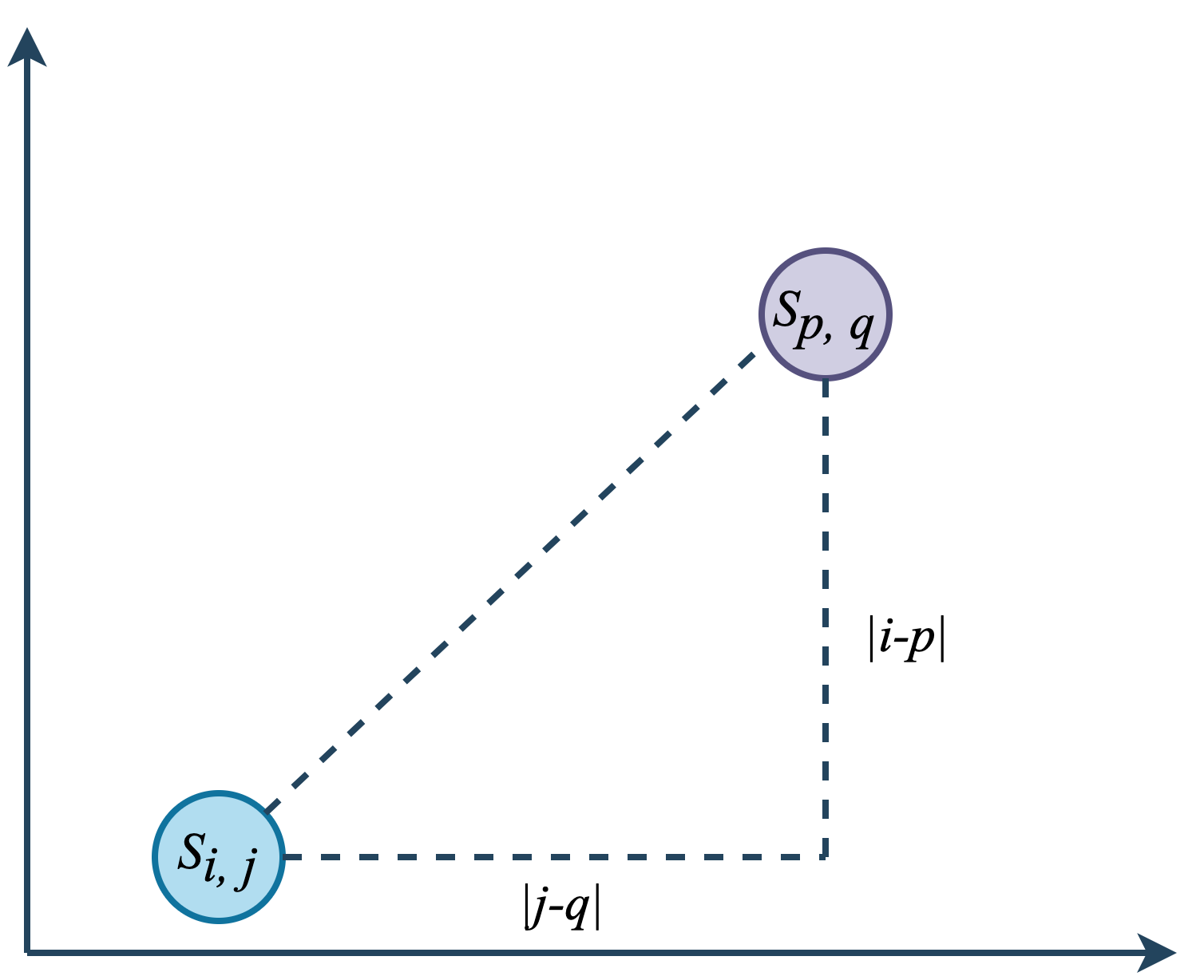}}
    \hfil
    \subfloat[]{\includegraphics[width=1.7in]{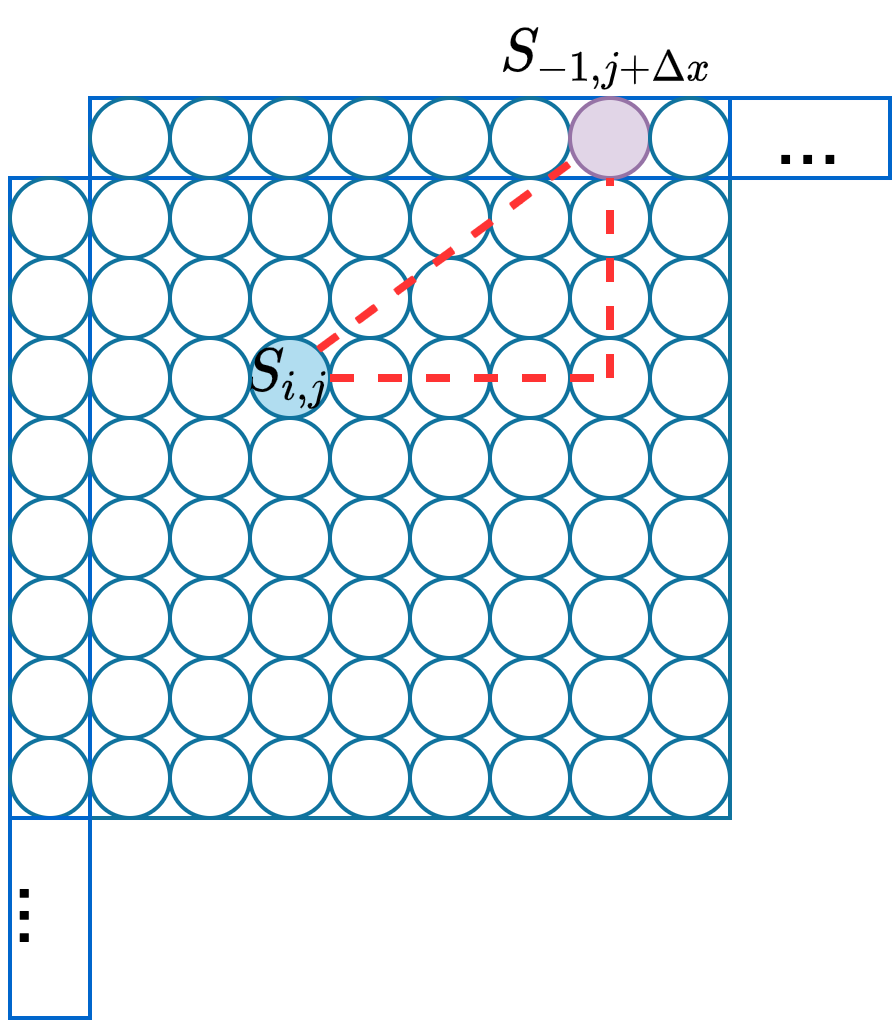}}
\caption{{Illustration of the inter-pixel correlations. (a) $S_{i,j}$ and $S_{p,q}$; (b) $S_{i,j}$ and $S_{-1,j+\Delta x}$.}}
\label{interPixel}
\end{figure}
\begin{figure}[t]
    \centering
    \subfloat[]{\includegraphics[width=2in]{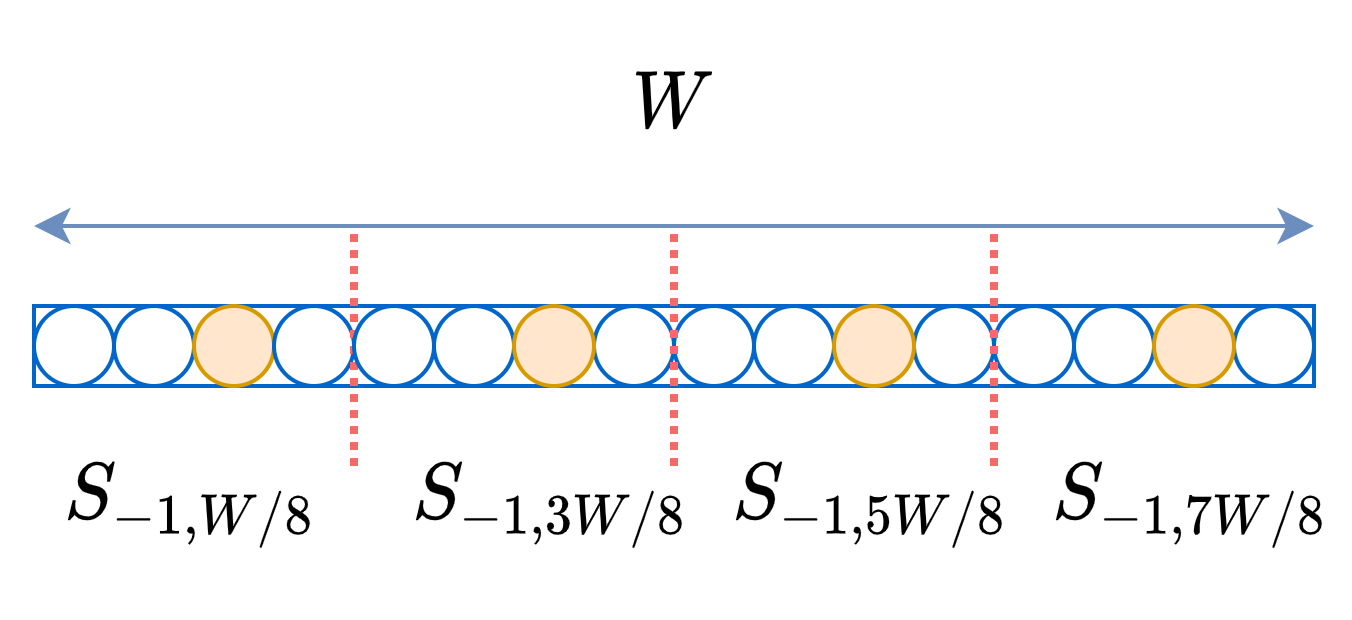}}
    \hfil
    \subfloat[]{\includegraphics[width=1.2in]{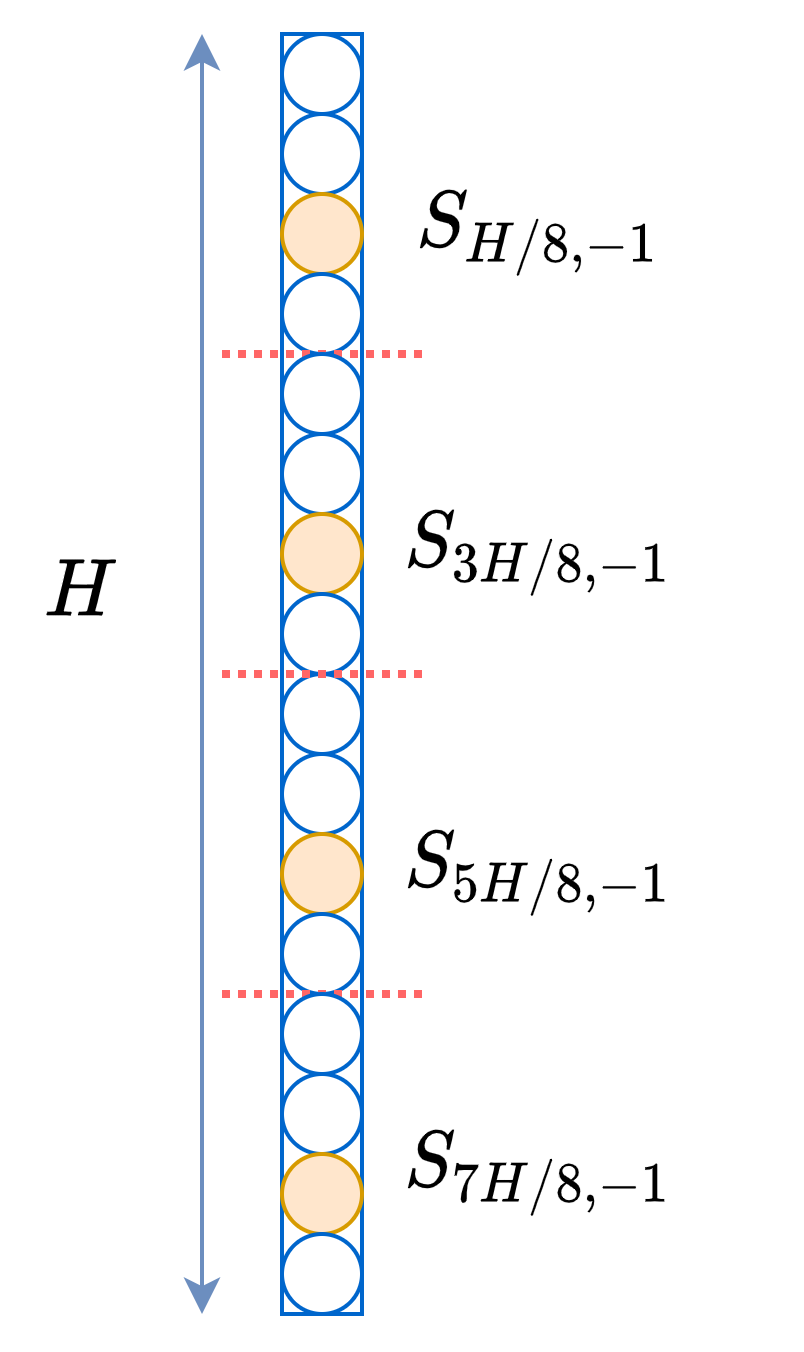}}
    \hfil
    \subfloat[]{\includegraphics[width=2.2in]{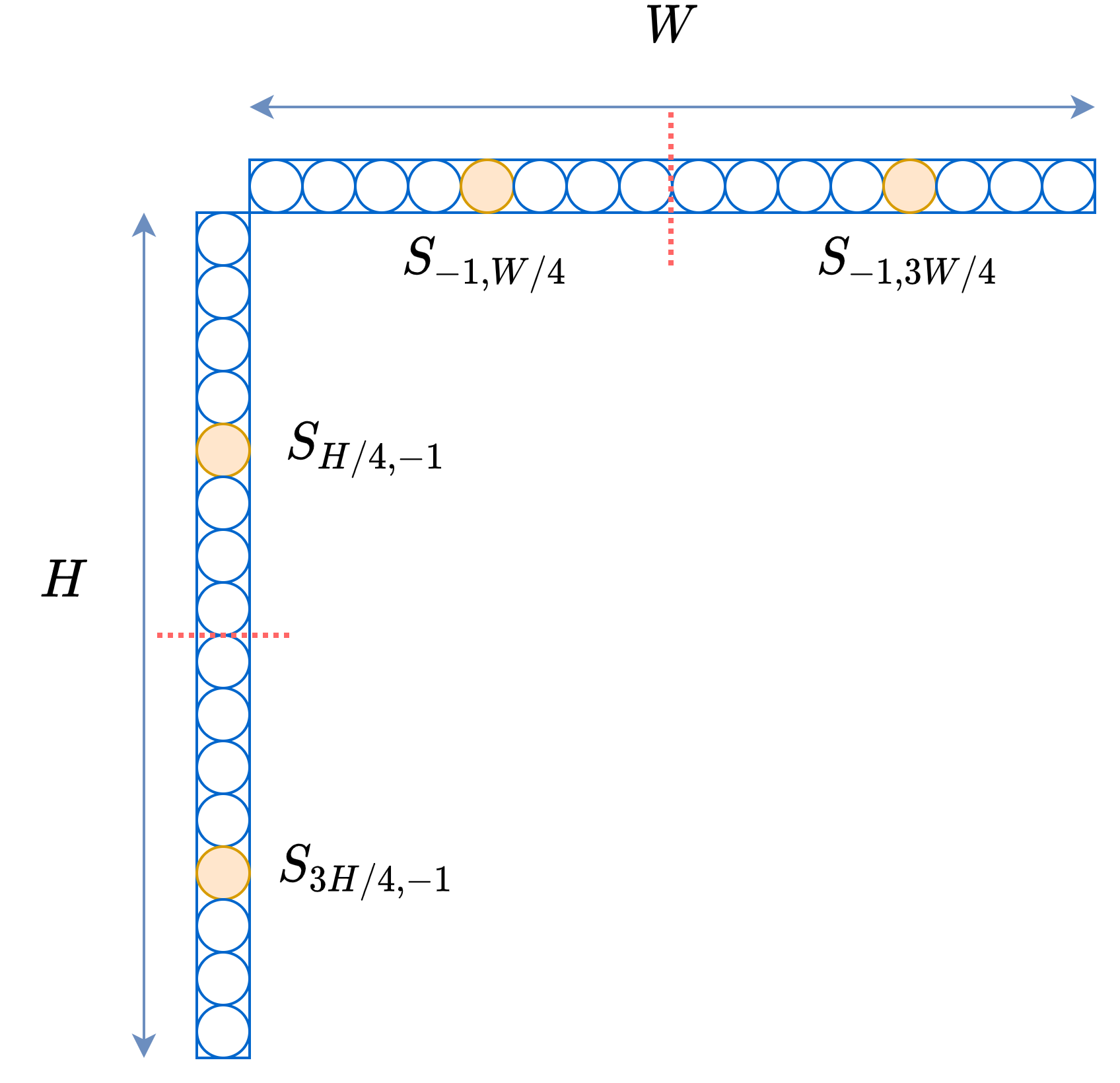}}
\caption{{Illustration of the position selections with different LM modes. (a) LM-Above mode; (b) LM-Left mode; (c) LM mode.}}
\label{SamplePosition}
\end{figure}
\begin{figure}[t]
    \centering
    {\includegraphics[width=2.2in]{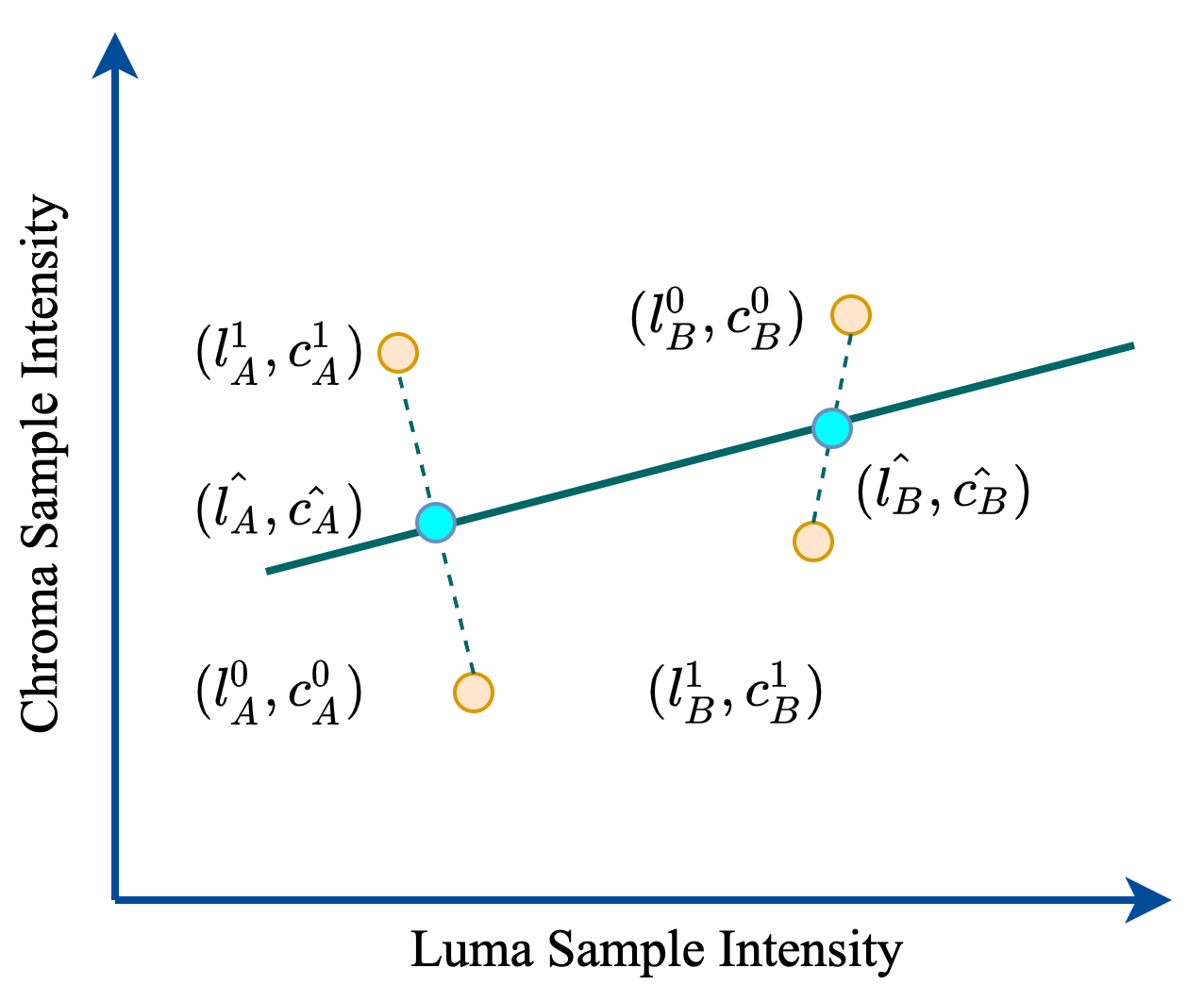}}
\caption{{Illustration of the smaller two samples and larger two samples.}}
\label{max2min2}
\end{figure}
\subsection{Selection of the Reference Sample Positions}
Based on the former analyses that only four reference sample pairs are involved in the model derivation, subsequently it is desirable to study which samples should be selected. It should be mentioned that the prerequisite of CCLM is grounded on the assumption that the inter-channel correlation of the reference samples is close to that of the current CB. Generally speaking, natural scene images and videos may contain various colors and texture details, such that the linear prediction with different color channels is essentially a simplified estimation. Pixels with higher correlations are more prone to share a similar model, leading to lower prediction errors, which inspires us to select the reference samples from the perspective of inter-pixel correlation. 

Denote $S_{i, j}$ and $S_{p, q}$ in Fig.~\ref{interPixel}(a) as two pixels locating at position $(i, j)$ and $(p, q)$, where the pixel is assumed to be a random variable with zero mean and unit variance. 
The correlation of $S_{i, j}$ and $S_{p, q}$ can be described according to the image correlation model~\cite{mauersberger1979generalised},
\begin{align}
    \rho(S_{i, j}, S_{p, q}) = \rho_x^{|j-q|}\cdot\rho_y^{|i-p|},
    \label{pixCorrelation}
\end{align}
where $\rho_x$ and $\rho_y$ denote the correlations of adjacent pixels along the horizontal and vertical directions, respectively.  
The pixel correlation model in Eqn.~(\ref{pixCorrelation}) suggests that besides the statistical characteristics regarding the adjacent pixels, the correlation of two pixels also relies on the distance between them. Moreover, when employing the reference sample $S_{-1, j+\Delta x}$ to predict the current sample $S_{i, j}$, as illustrated in Fig.~\ref{interPixel}(b), the prediction error associated to sample $S_{i, j}$ can be formulated as follows,
\begin{align}
    e_{i, j} = S_{i, j} - S_{-1, j+\Delta x}.
    \label{predErr}
\end{align}
By combining Eqn.~(\ref{pixCorrelation}) and Eqn.~(\ref{predErr}), we can obtain the co-variance of the prediction error between $e_{i, j}$ and $e_{p, q}$ as follows,
\begin{align}
    &cov(e_{i, j}, e_{p, q}) \nonumber \\
    &= E[(e_{i, j}-\mu_e)(e_{p, q}-\mu_e)] \nonumber \\
    &= E[(S_{i, j} - S_{-1, j+\Delta x})(S_{p, q} - S_{-1, q+\Delta x})] \nonumber\\
    &= E[S_{i, j}S_{p, q}]-E[S_{-1, j+\Delta x}S_{p, q}] \nonumber \\
    &- E[(S_{i, j}S_{-1, q+\Delta x}]+E[S_{-1, j+\Delta x}S_{-1, q+\Delta x}]\nonumber\\
    &= \rho_x^{|j-q|}\rho_y^{|i-p|} - \rho_x^{|j+\Delta x-q|}\rho_y^{|-1-p|}-\rho_x^{|j-q-\Delta x|}\rho_y^{|i+1|} + \rho_x^{j-q},
\end{align}
where $\mu_e$ represents the expectation of the prediction error which is assumed to be zero. A common criterion to evaluate the prediction efficiency is the variance of the prediction error~\cite{AngularIntra_TIP2014}. By substituting the $p$ and $q$ with $i$ and $j$, the variance of the prediction error $\sigma_{i,j}^2$ can be described as follows,
\begin{align}
    \sigma_{i,j}^2 = 2 - 2\rho_y^{i+1}\rho_x^{|\Delta x|}.
\end{align}
As such, $\sigma_{i,j}^2$ decreases with the decreasing of $i$ and $|\Delta x|$.

Herein, $\sigma_{i,j}^2$ typically depicts the prediction efficiency of a single pixel. When considering the total variance of the prediction error for a row of samples with respect to one reference sample at position $(-1, t)$, the problem can be formulated as follows,
\begin{align}
    \sum_{j=0}^{N-1}\sigma_{i,j}^2 = 2N-2\rho_y^{i+1}\cdot f(t),
\end{align}
where
\begin{align}
    f(t) = \sum_{j=0}^{N-1}\rho_x^{|j-t|}
    =\frac{1+\rho_x-\rho_x^{t+1}-\rho_x^{N-t}}{1-\rho_x}.
\end{align}
It is imperative to explore the minimum value regarding the total variance of the prediction error, which corresponds to the high prediction efficiency. In particular, minimizing $\sum_{j=0}^{N-1}\sigma_{i,j}^2$ is equivalent to maximizing $f(t)$. By setting the deviation of $f(t)$ with respect to $t$ to zero, we have,
\begin{align}
    f'(t)=\frac{-(\rho_x^{t+1}-\rho_x^{N-t})\cdot \ln\rho_x}{1-\rho_x}=0.
\end{align}
As such, we can obtain the extreme maximum value of $f(t)$ at $t_0$, where $t_0 = (N-1)/2$. In this sense, we can conclude that the sample located at middle position achieves better prediction efficiency. 

Therefore, in our method, the set of reference samples will be equally divided into two or four parts depending on the explicit mode, and the sample pairs with middle positions will be involved in the linear model derivation, as illustrated in Fig.~\ref{SamplePosition}. Supposing there are $W$ and $H$ reference sample pairs in the above and left side, respectively, when the LM-Above mode is employed, the above reference line will be equally divided into four parts and four middle sample within each part $(S_{-1, W/8}, S_{-1, 3W/8}, S_{-1, 5W/8}, S_{-1, 7W/8})$ are selected. Reference samples with middle positions possess relatively higher correlations with the current predicted block. Furthermore, considering the fact that the regression dilution problem is quite common in linear regression, which may cause biased estimation and impair the prediction efficiency, the proposed separation strategy tries to ensure the diversification of the selected sample pairs such that the regression dilution can be averted.

\subsection{Linear Model Derivation}
Instead of employing the entire reference line or column of sample pairs to derive the linear model, with the proposed method, at most four sample pairs with relative fixed positions are utilized. More specifically, the sub-sampled pairs can be feasibly picked with predefined offsets when encountering different dimensions of reference sample sets. In this manner, luma down-sampling is only applied to the specific samples, which is conducive to reduce the operation complexity. To improve the robustness of the linear model, the four sample pairs are grouped into two larger and two smaller couples based on luma intensities. Subsequently, the smaller two samples $(l_A^0, c_A^0)$, $(l_A^1, c_A^1)$ and larger two samples $(l_B^0, c_B^0)$, $(l_B^1, c_B^1)$ as illustrated in Fig.~\ref{max2min2}, are averaged as follows,
\begin{align}
    \hat{l_A} &= \frac{l_A^0+l_A^1}{2},
    \hat{l_B} = \frac{l_B^0+l_B^1}{2}, \nonumber \\
    \hat{c_A} &= \frac{c_A^0+c_A^1}{2},
    \hat{c_B} = \frac{c_B^0+c_B^1}{2}.
\end{align}
The model parameters $\alpha$ and $\beta$ can be derived as,
\begin{align}
    \alpha &= \frac{\hat{c_B} - \hat{c_A}}{\hat{l_B} - \hat{l_A}}, \nonumber \\
    \beta  &= \hat{c_A} - \alpha \cdot \hat{l_A}.
    \label{max2min2alphabeta}
\end{align}
For the case that only two reference sample pairs are accessible, such as the 2 $\times$ 2 chroma CB predicted with LM-Above or LM-Left mode, the model parameters are derived with the two available samples. 

\subsection{Analyses on the Sensitivity of the Model Parameters}
Furthermore, we investigate the sensitivity of the model parameters when cooperating the proposed scheme. Given the luma sample set $L$ and the corresponding chroma sample set $C$, the linear model parameters derived through LSR are regarded as the benchmark. The deformation of $\alpha_{LSR}$ and $\beta_{LSR}$ are described as follows,
\begin{align}
    \alpha_{LSR} &= \frac{cov(L,C)}{\sigma_L^2},\nonumber\\
    \beta_{LSR} &= \mu_C - \mu_L\cdot\alpha_{LSR}.
    \label{alpha_beta_cov}
\end{align}
The associated prediction efficiency can be measured by the variance of the prediction error as follows,
\begin{align}
    e(\alpha_{LSR}, \beta_{LSR}) &= E[(C -\alpha_{LSR}\cdot L - \beta_{LSR} - \mu_{r})^2] \nonumber \\
    &= E[\left((C - \mu_C) - \alpha_{LSR}(L - \mu_L)\right)^2] \nonumber \\
    &= \sigma_C^2 + \alpha_{LSR}^2\sigma_L^2 - 2\alpha_{LSR} \cdot cov(L, C) \nonumber \\
    &= \sigma_C^2 - \frac {cov(L,C)^2} {\sigma_L^2}.
\end{align}
Herein, $\mu_r$ represents the mean of the residuals which is assumed to be zero.

When using the proposed method or Max-Min method to derive the linear model parameters, offsets $\Delta\alpha$ and $\Delta\beta$ are imposed to the $\alpha_{LSR}$ and $\beta_{LSR}$ as follows,
\begin{align}
    \hat{\alpha} &= \alpha_{LSR} + \Delta\alpha, \nonumber \\
    \hat{\beta}  &= \beta_{LSR} + \Delta\beta.
\end{align}
In this manner, the associated $e(\hat{\alpha}, \hat{\beta})$ can be formulated as,
\begin{align}
    e(\hat{\alpha}, \hat{\beta}) &= E[(C -(\alpha_{LSR}+\Delta\alpha)\cdot L - (\beta_{LSR}+\Delta\beta) - \mu_{r}^{'})^2] \nonumber\\
    &= E[((C -\alpha_{LSR}\cdot L -\beta_{LSR}) - (\Delta\alpha \cdot L + \Delta\beta))^2] \nonumber\\
    &= E[(C -\alpha_{LSR}\cdot L -\beta_{LSR})^2] + \gamma + 2 \cdot \eta \nonumber\\
    &= \sigma_C^2 - \frac {cov(L,C)^2} {\sigma_L^2} + \gamma + 2 \cdot \eta,
\end{align}
where
\begin{align}
    \gamma = E[(\Delta\alpha \cdot L + \Delta\beta)^2],
\end{align}
and
\begin{align}
    \eta = E[(\alpha_{LSR}\cdot L + \beta_{LSR} - C)(\Delta\alpha \cdot L + \Delta\beta)].
\end{align}
According to Eqn.~(\ref{alpha_beta_cov}), $\Delta \beta$ can be further written as -$\Delta\alpha\cdot\mu_L$. As such, $\gamma$ and $\eta$ can be simplified as follows,
\begin{align}
    \gamma &= E[(\Delta \alpha \cdot L - \Delta\alpha\cdot\mu_L)^2] \nonumber\\
    &= \Delta\alpha^2 \cdot E[(L - \mu_L)^2] \nonumber\\
    &= \Delta\alpha^2 \cdot \sigma_L^2, \nonumber \\
    \eta &= E[(\alpha_{LSR}\cdot L + \beta_{LSR} - C)(\Delta\alpha \cdot L - \Delta\alpha \cdot \mu_L)] \nonumber\\
    &= \Delta\alpha \cdot E[ (\alpha_{LSR}\cdot (L - \mu_L) - (C - \mu_C)) (L - \mu_L) ] \nonumber\\
    &= \Delta\alpha \cdot \{\alpha_{LSR}\cdot E[(L - \mu_L)^2] - E[(C - \mu_C) \cdot (L - \mu_L)]\} \nonumber\\
    &= \Delta\alpha \cdot (\alpha_{LSR} \cdot \sigma_{L}^2 - cov(L,C)) \nonumber\\
    &= 0.
\end{align}
As such, the variance of prediction error is increased by $\Delta e$,
\begin{align}
    \Delta e &= e(\hat{\alpha}, \hat{\beta}) - e(\alpha_{LSR}, \beta_{LSR}) \nonumber\\
    &= \Delta\alpha^2\cdot \sigma_L^2.
    \label{DeltaE}
\end{align}

\begin{figure}[t]
    \centering
    \subfloat[]{\includegraphics[width=1.7in]{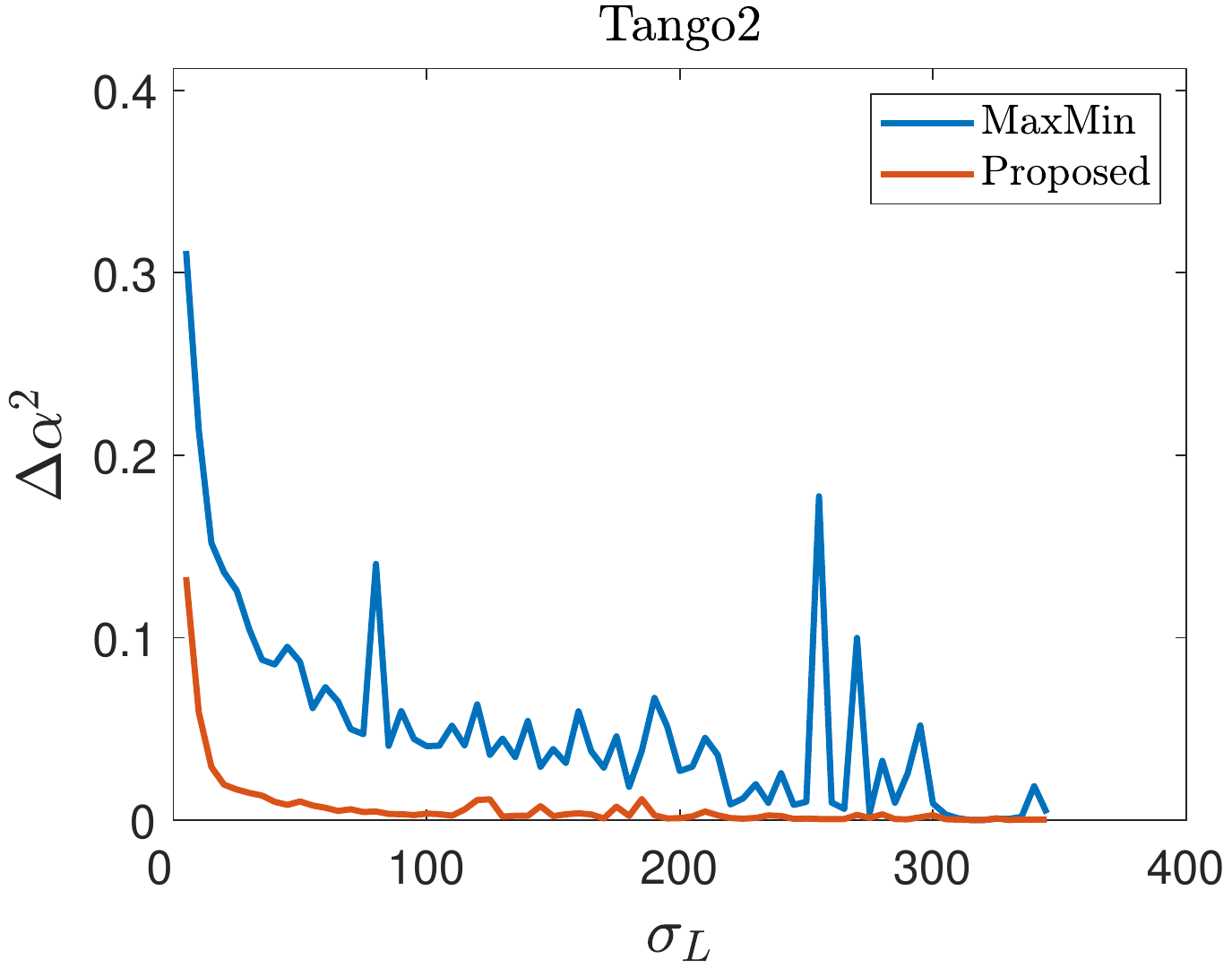}}
    \hfil
    \subfloat[]{\includegraphics[width=1.7in]{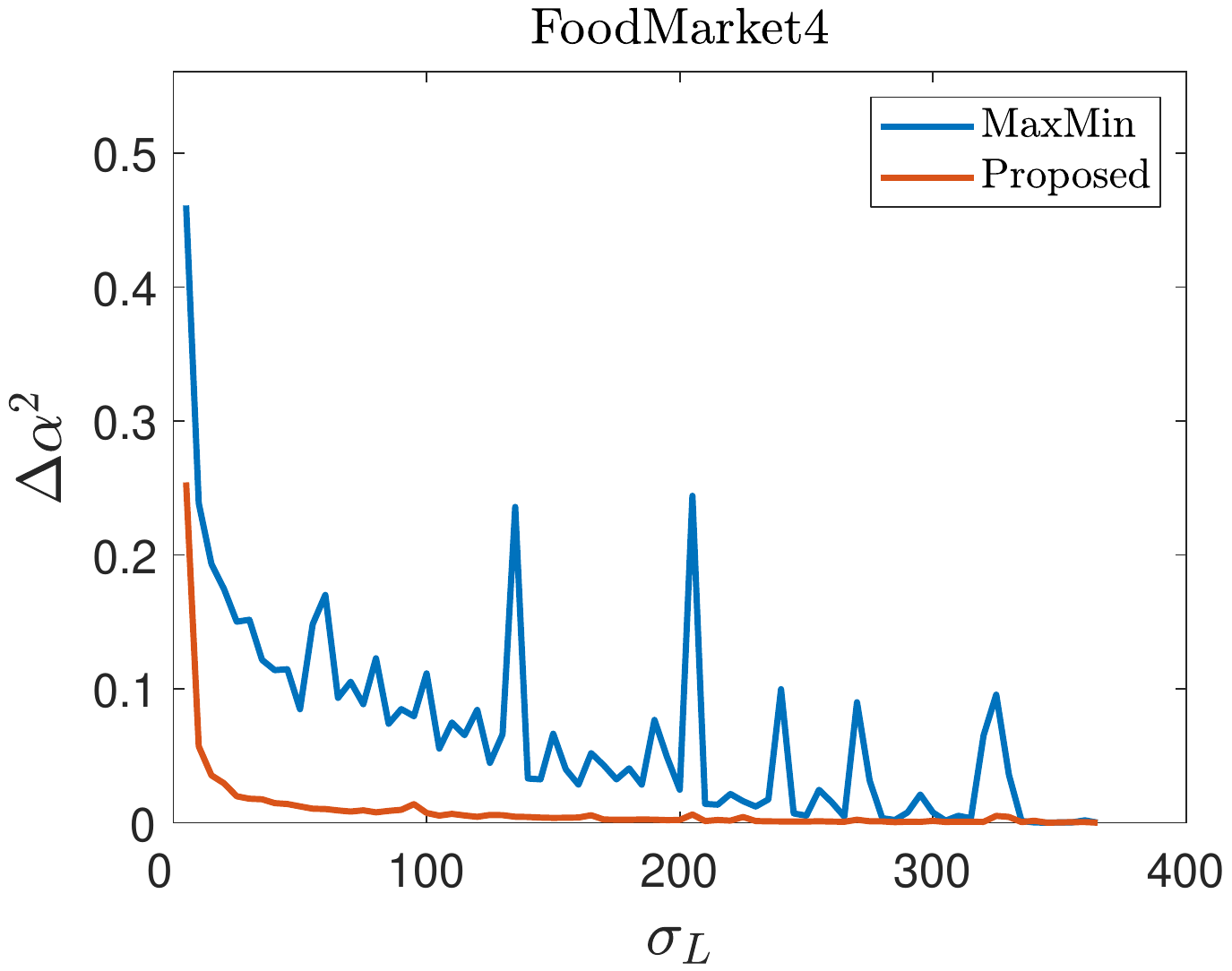}}\\
    
    \subfloat[]{\includegraphics[width=1.7in]{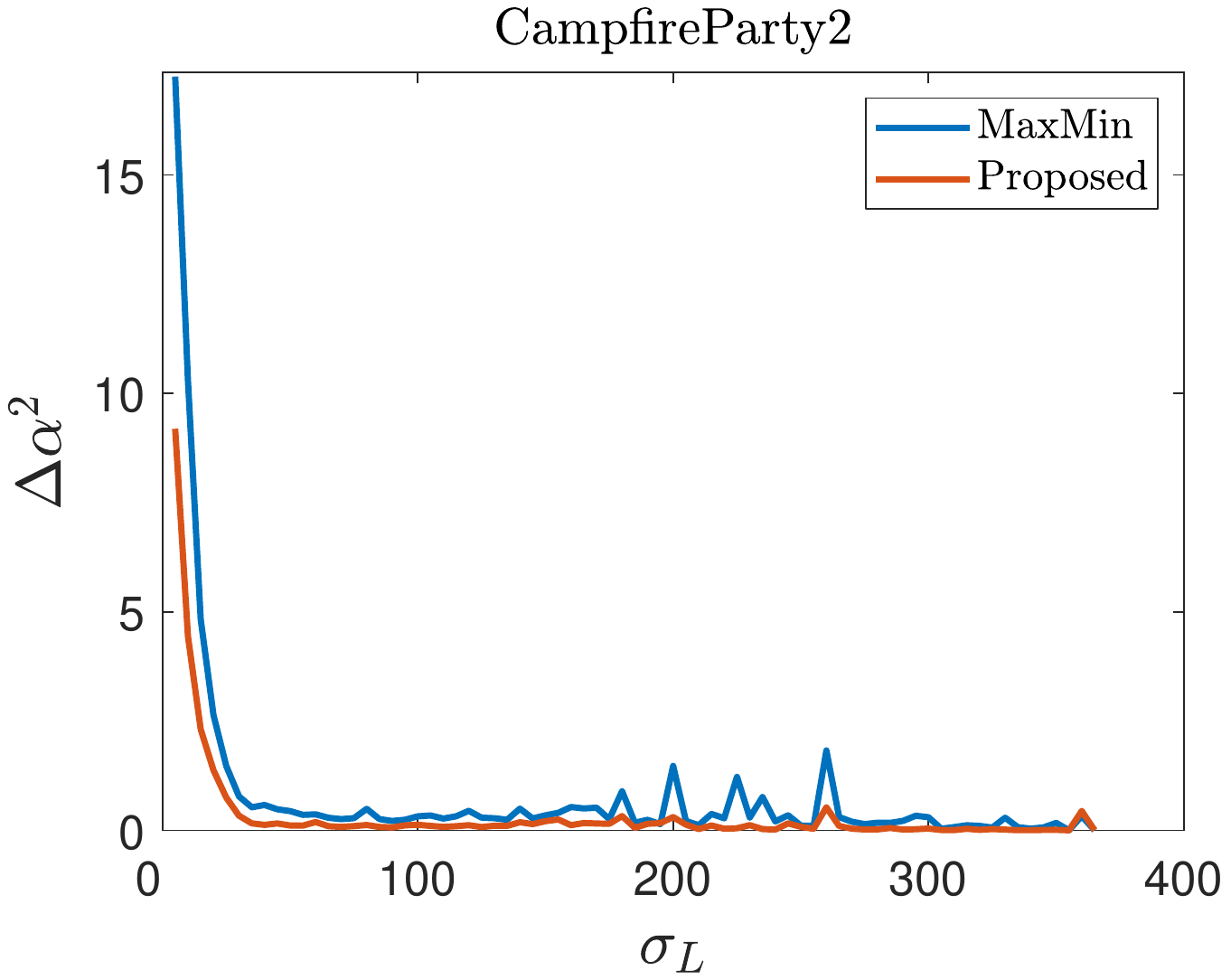}}
    \hfil
    \subfloat[]{\includegraphics[width=1.7in]{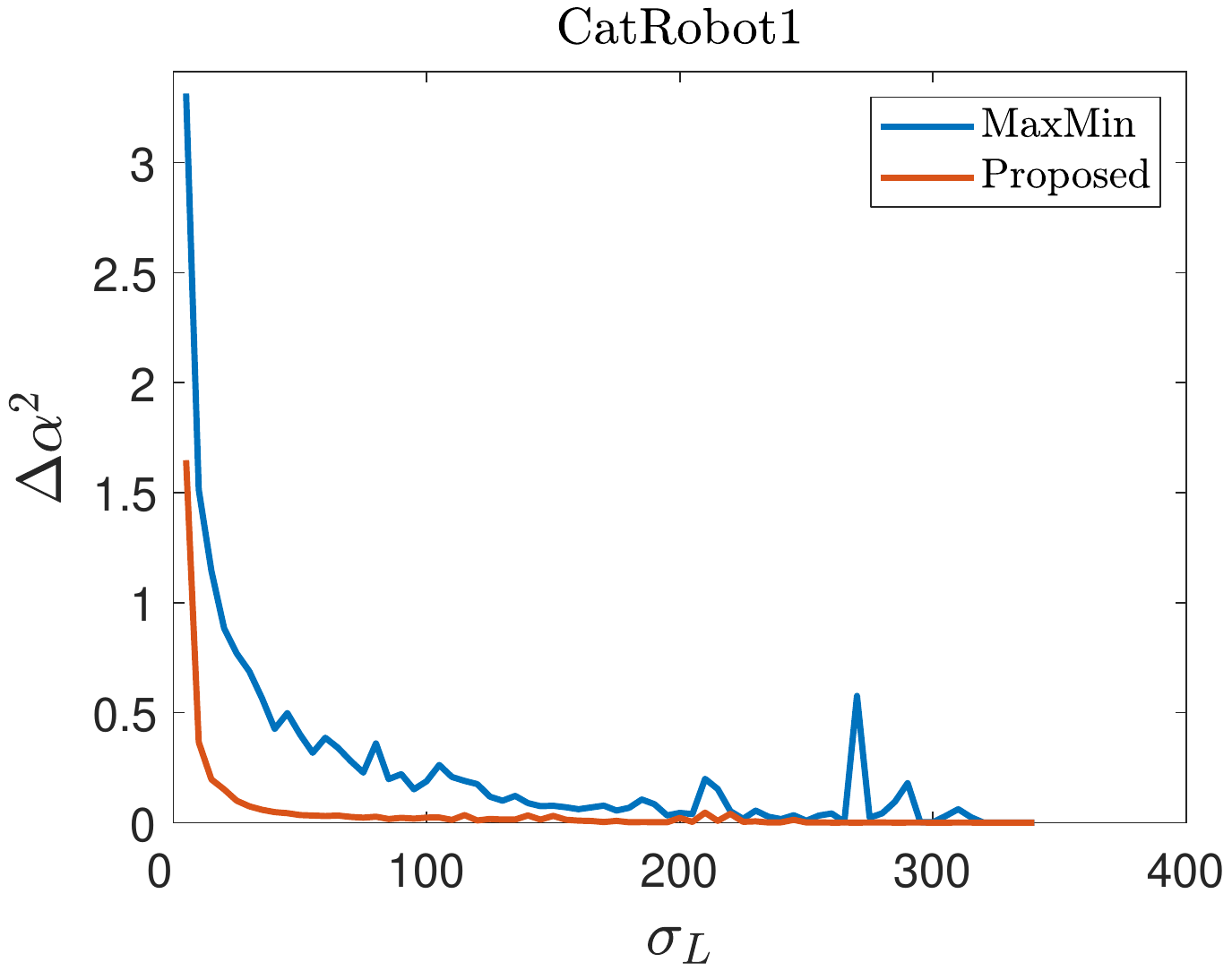}}\\
\caption{{Illustration of $\Delta \alpha^2$ with regard to the proposed derivation method and Max-Min method.}}
\label{deltaAlpha}
\end{figure}

\begin{figure*}[t]
    \centering
    \subfloat[]{\includegraphics[width=1.7in]{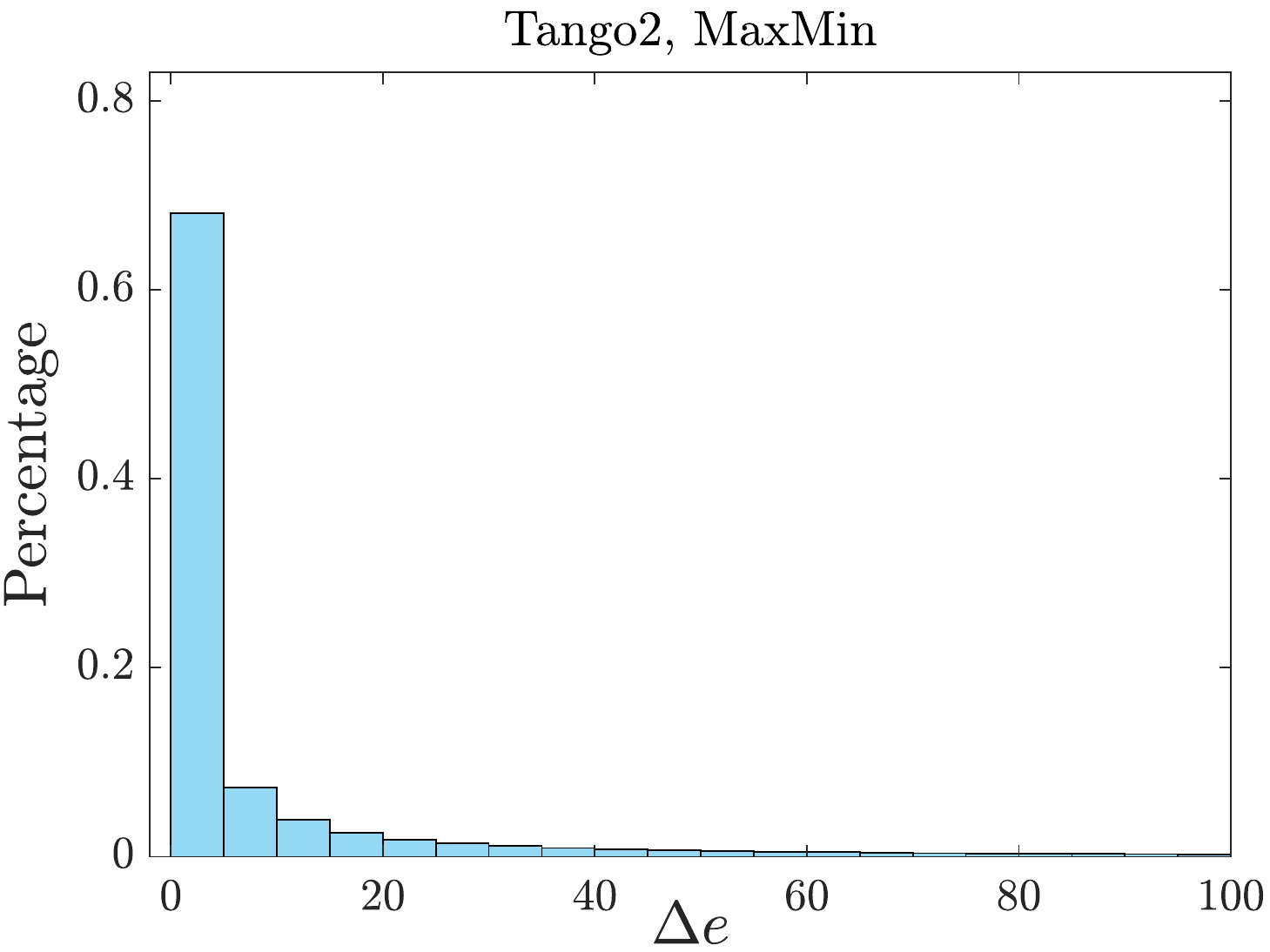}}
    \hfil
    \subfloat[]{\includegraphics[width=1.7in]{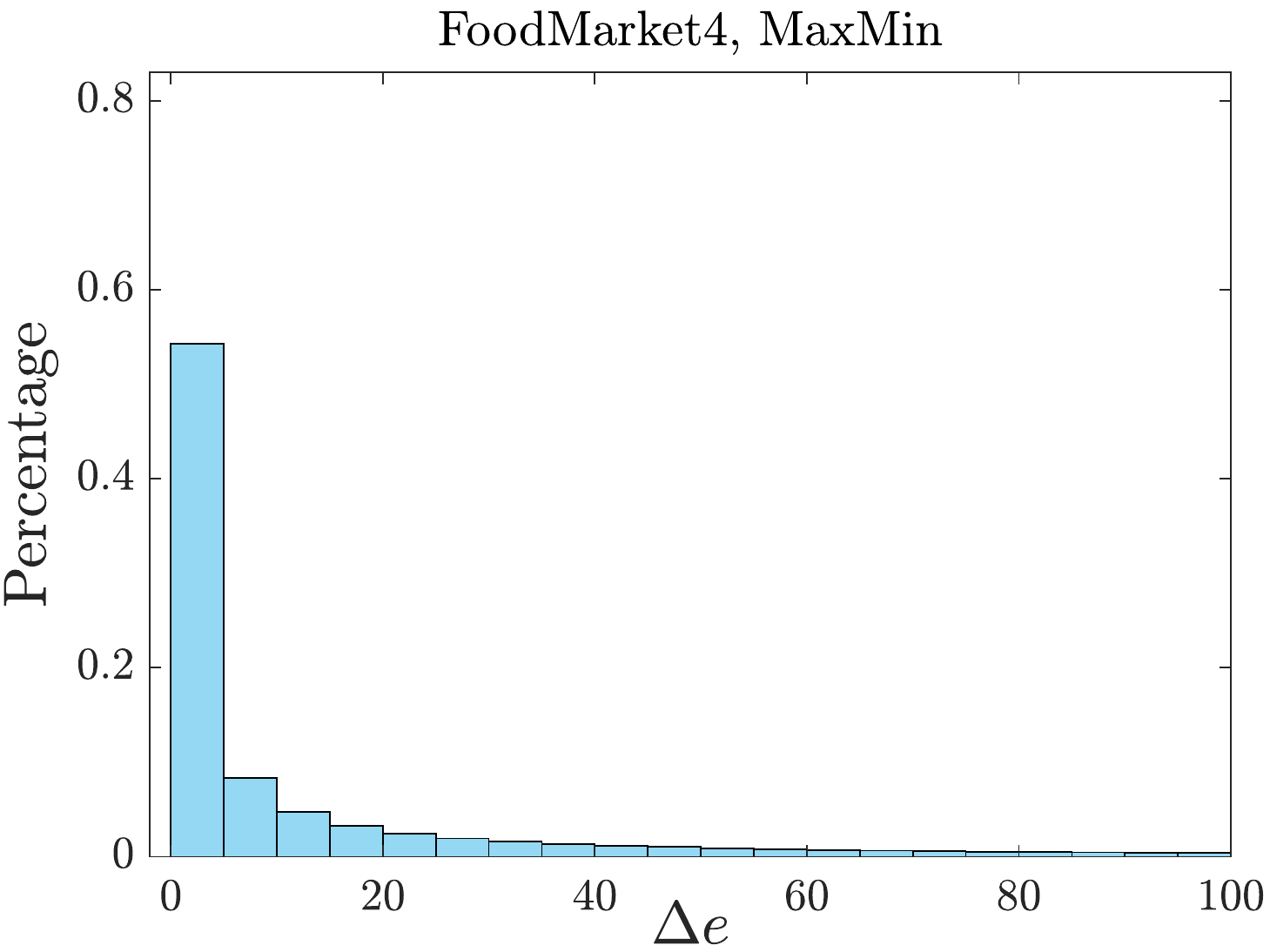}}
    \hfil
    \subfloat[]{\includegraphics[width=1.7in]{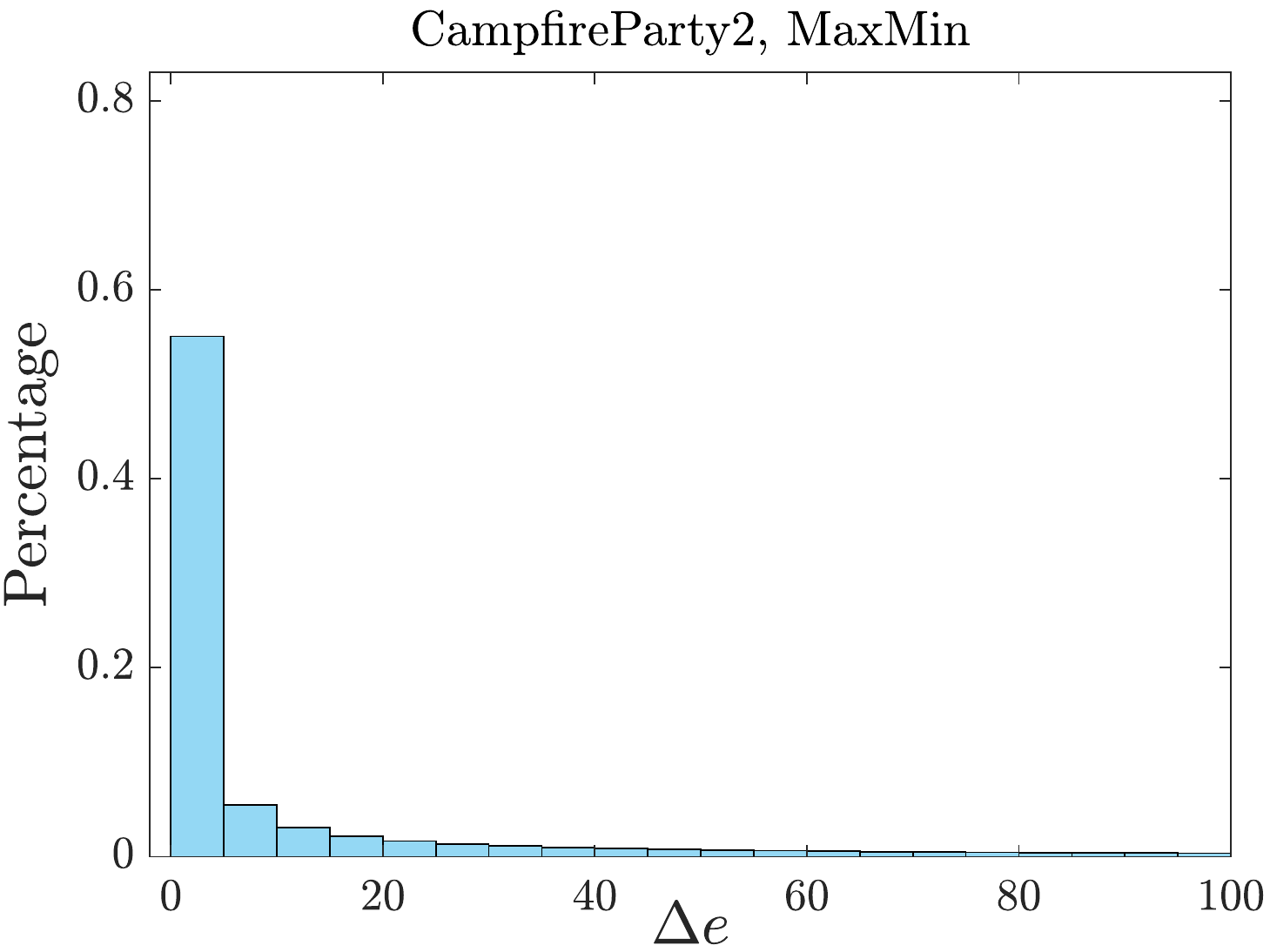}}
    \hfil
    \subfloat[]{\includegraphics[width=1.7in]{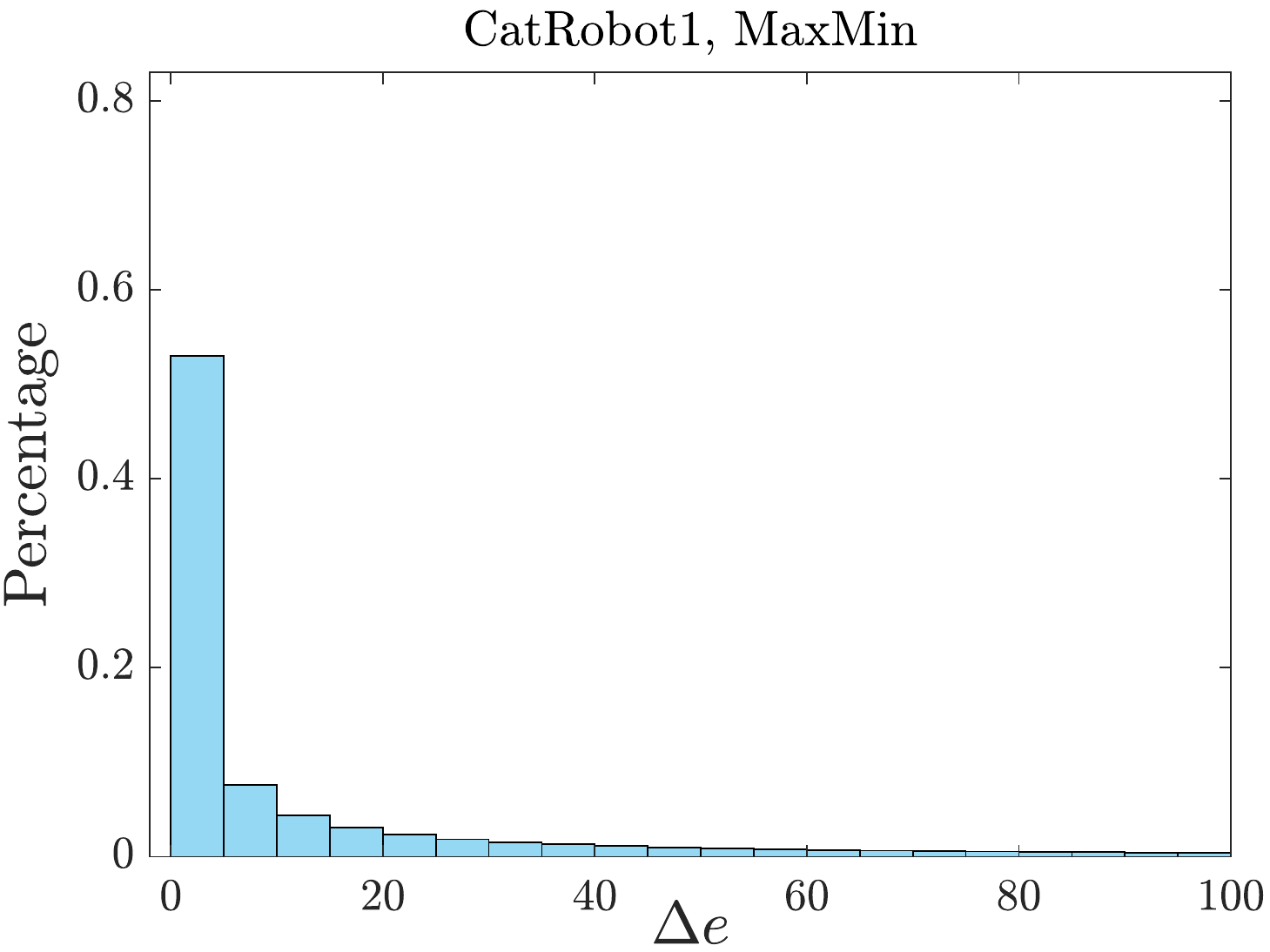}}\\
	
	\subfloat[]{\includegraphics[width=1.7in]{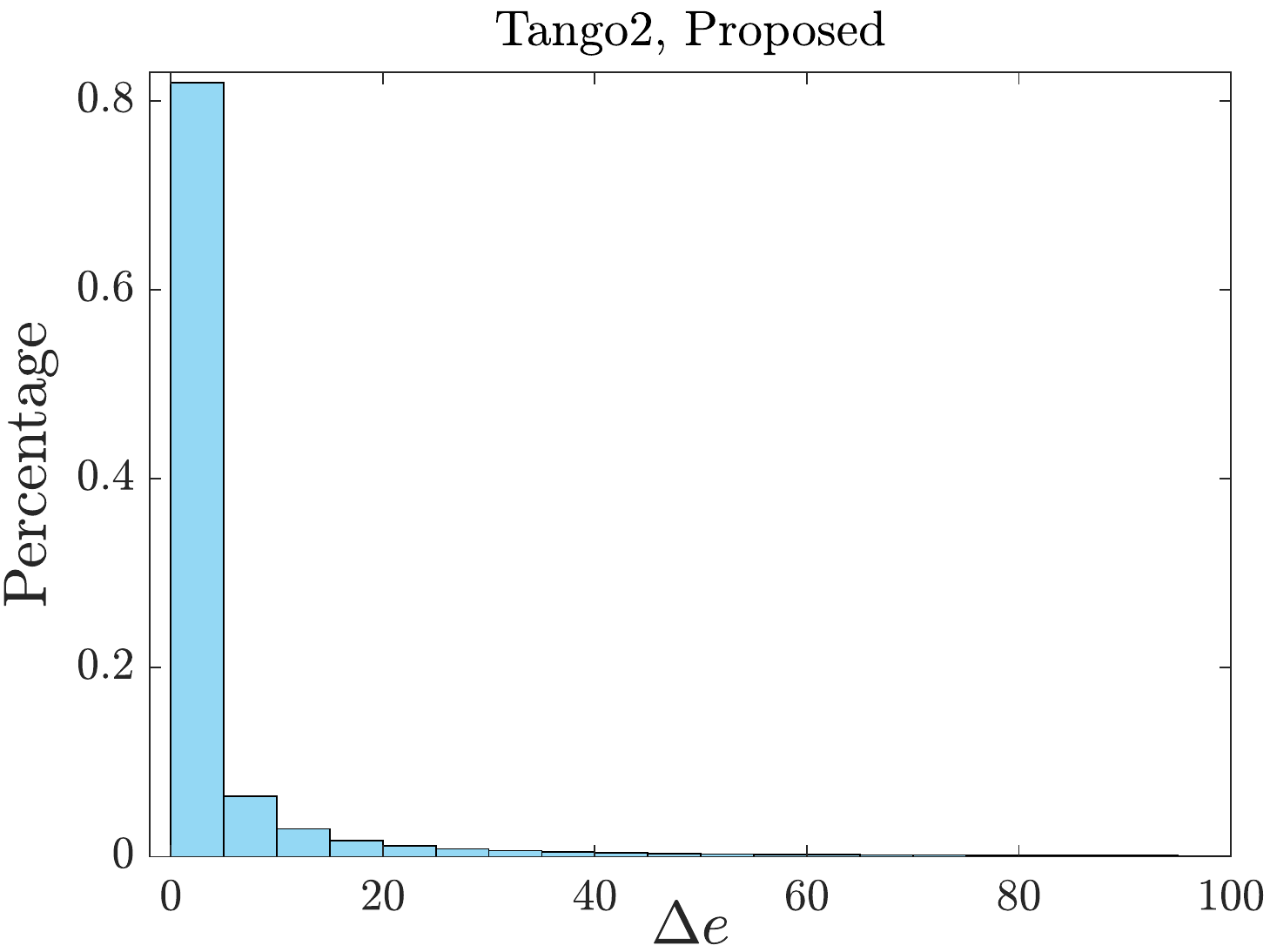}}
    \hfil
    \subfloat[]{\includegraphics[width=1.7in]{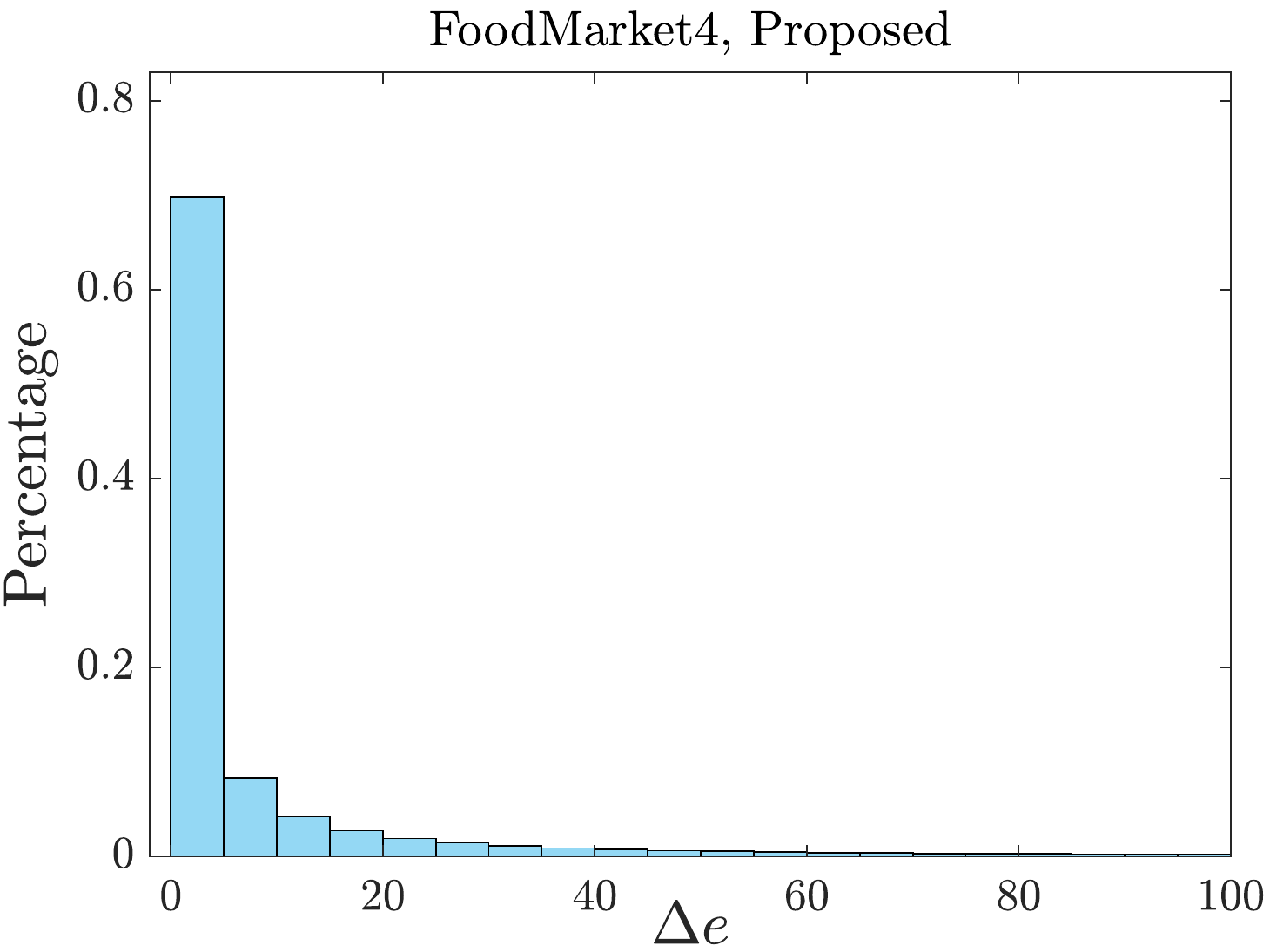}}
    \hfil
    \subfloat[]{\includegraphics[width=1.7in]{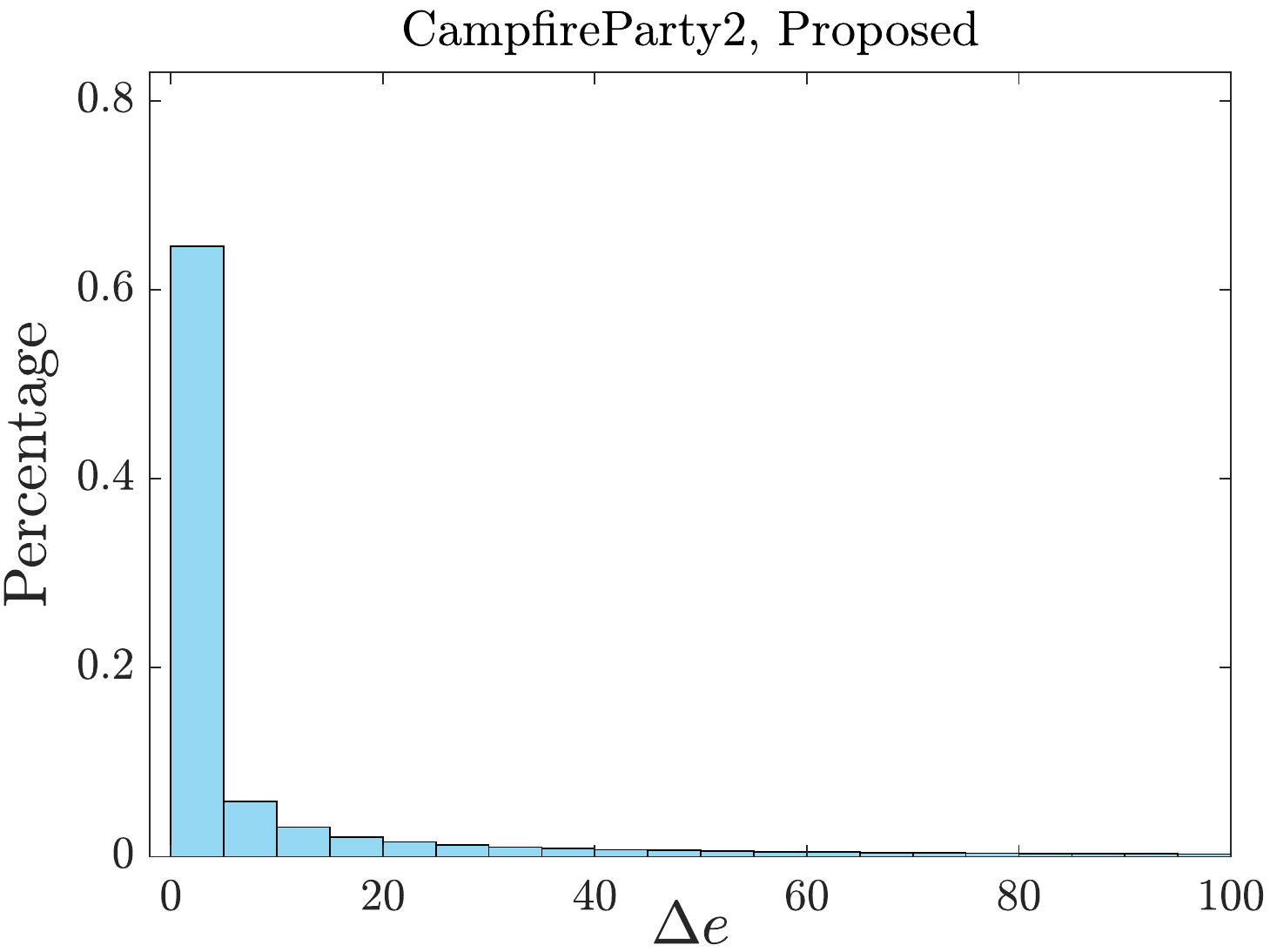}}
    \hfil
    \subfloat[]{\includegraphics[width=1.7in]{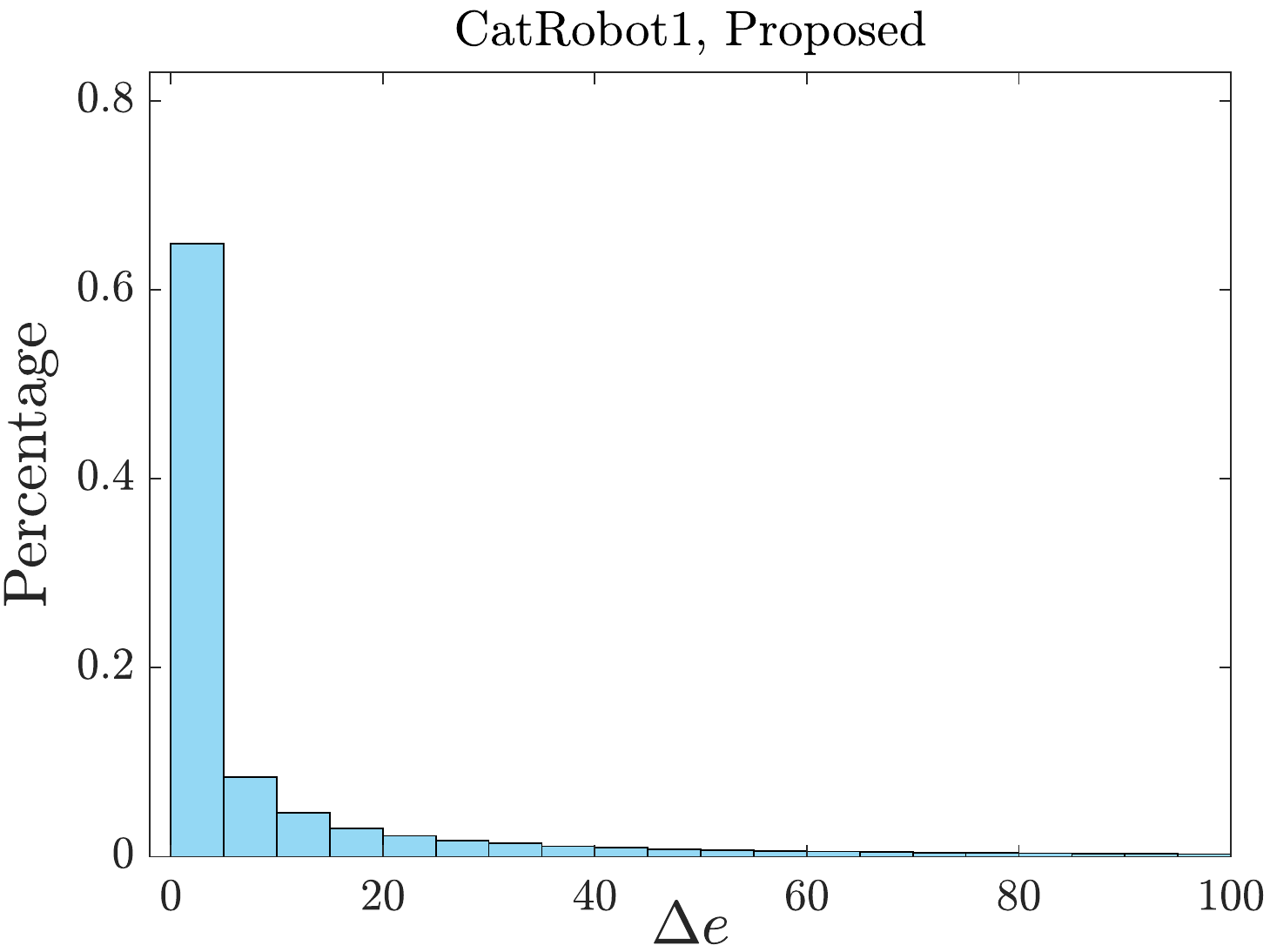}}\\
	
\caption{{Illustration of the percentage of $\Delta e$ with regard to the proposed derivation method and Max-Min method.}}
\label{deltaE}
\end{figure*}

It can be noticed that $\Delta e$ increases with the variation of luma component $\sigma_L^2$ and the $\Delta \alpha^2$. For specific video content where $\sigma_L^2$ is unchanged, $\Delta \alpha^2$ can reflect the robustness of the linear derivation model. Higher $\Delta \alpha^2$ may magnify the sensitivity of the model parameters. We statistically analyze $\Delta \alpha^2$ with regard to the proposed derivation method and Max-Min method, and the results are illustrated in Fig.~\ref{deltaAlpha}. Here, four test sequences ``Tango2'', ``FoodMarket4'', ``CampfireParty2'' and ``CatRobot1'' are involved . We can observe that $\Delta \alpha^2$ of the proposed method is constantly lower than that of Max-Min method with varied {$\sigma_L$}. Moreover, the percentage of $\Delta e$ is illustrated in Fig.~\ref{deltaE}, and $\Delta e$ of the proposed method is more centralized with zero.

\begin{table*}[htbp]
  \centering
  \caption{Coding Performance of the Proposed Scheme Compared with Max-Min Scheme on VTM-5.0 Testing Platform under AI, RA and LDB Configurations}
    \begin{tabular}{cc|ccc|ccc|ccc}
    \toprule
    \multicolumn{2}{c|}{\multirow{2}[2]{*}{Class}} & \multicolumn{3}{c|}{AI} & \multicolumn{3}{c|}{RA} & \multicolumn{3}{c}{LDB} \\
    \cmidrule{3-11} \multicolumn{2}{c|}{} & Y & U & V & Y & U & V & Y & U & V \\
    \midrule
    \multicolumn{2}{c|}{A1} & -0.20\% & 0.07\% & -0.34\% & -0.19\% & 0.82\% & -0.53\% & -     & -     & - \\
    \multicolumn{2}{c|}{A2} & -0.03\% & -0.14\% & 0.00\% & -0.02\% & -0.14\% & 0.00\% & -     & -     & - \\
    \multicolumn{2}{c|}{B} & -0.02\% & -0.48\% & -0.62\% & -0.02\% & -0.72\% & -0.58\% & -0.02\% & -0.46\% & -0.33\% \\
    \multicolumn{2}{c|}{C} & -0.05\% & -0.33\% & -0.35\% & 0.01\% & -0.08\% & -0.11\% & 0.02\% & -0.35\% & -0.27\% \\
    \multicolumn{2}{c|}{E} & -0.02\% & -0.04\% & 0.08\% & -     & -     & -     & -0.07\% & 0.35\% & -1.45\% \\
    \midrule
    \multicolumn{2}{c|}{\textbf{Overall}} & \textbf{-0.06\%} & \textbf{-0.23\%} & \textbf{-0.29\%} & \textbf{-0.05\%} & \textbf{-0.13\%} & \textbf{-0.33\%} & \textbf{-0.02\%} & \textbf{-0.22\%} & \textbf{-0.59\%} \\
    \midrule
    \multicolumn{2}{c|}{D} & -0.01\% & -0.25\% & -0.02\% & 0.00\% & -0.25\% & -0.32\% & -0.04\% & 0.57\% & -1.02\% \\
    \multicolumn{2}{c|}{F} & -0.04\% & -0.18\% & 0.02\% & 0.03\% & 0.08\% & -0.16\% & -0.09\% & 0.21\% & 0.58\% \\
    \midrule
    \multicolumn{2}{c|}{Enc Time} & \multicolumn{3}{c|}{99\%} & \multicolumn{3}{c|}{99\%} & \multicolumn{3}{c}{100\%} \\
    \midrule
    \multicolumn{2}{c|}{Dec Time} & \multicolumn{3}{c|}{100\%} & \multicolumn{3}{c|}{99\%} & \multicolumn{3}{c}{99\%} \\
    \bottomrule
    \end{tabular}%
  \label{Perf_VVC}%
\end{table*}%

\begin{table*}[htbp]
  \centering
  \caption{  Coding Performance of CCLM on VTM-5.0 Testing Platform under AI, RA and LDB Configurations}
    \begin{tabular}{cc|ccc|ccc|ccc}
    \toprule
    \multicolumn{2}{c|}{\multirow{2}[2]{*}{Class}} & \multicolumn{3}{c|}{AI} & \multicolumn{3}{c|}{RA} & \multicolumn{3}{c}{LDB} \\
    \cmidrule{3-11} \multicolumn{2}{c|}{} & Y & U & V & Y & U & V & Y & U & V \\
    \midrule
    \multicolumn{2}{c|}{A1} & -4.97\% & -28.03\% & -27.90\% & -2.60\% & -19.38\% & -24.40\% & -     & -     & - \\
    \multicolumn{2}{c|}{A2} & -1.94\% & -13.06\% & -7.40\% & -0.65\% & -12.05\% & -5.82\% & -     & -     & - \\
    \multicolumn{2}{c|}{B} & -0.84\% & -10.74\% & -15.42\% & -0.27\% & -12.81\% & -14.99\% & -0.05\% & -5.41\% & -6.59\% \\
    \multicolumn{2}{c|}{C} & -1.54\% & -12.55\% & -12.92\% & -0.48\% & -10.03\% & -9.67\% & -0.15\% & -3.91\% & -3.62\% \\
    \multicolumn{2}{c|}{E} & -0.22\% & -4.07\% & -4.73\% &   -    &    -   &   -    & -0.19\% & -2.91\% & -5.26\% \\
    \midrule
    \multicolumn{2}{c|}{\textbf{Overall}} & \textbf{-1.76\%} & \textbf{-13.30\%} & \textbf{-13.83\%} & \textbf{-0.87\%} & \textbf{-13.23\%} & \textbf{-13.62\%} & \textbf{-0.12\%} & \textbf{-4.28\%} & \textbf{-5.27\%} \\
    \midrule
    \multicolumn{2}{c|}{D} & -1.00\% & -10.43\% & -9.55\% & -0.35\% & -9.38\% & -8.49\% & -0.08\% & -3.33\% & -3.71\% \\
    \multicolumn{2}{c|}{F} & -3.27\% & -13.99\% & -16.00\% & -1.90\% & -10.78\% & -11.84\% & -0.63\% & -5.40\% & -5.96\% \\
    \midrule
    \multicolumn{2}{c|}{Enc Time} & \multicolumn{3}{c|}{100\%} & \multicolumn{3}{c|}{101\%} & \multicolumn{3}{c}{100\%} \\
    \midrule
    \multicolumn{2}{c|}{Dec Time} & \multicolumn{3}{c|}{101\%} & \multicolumn{3}{c|}{99\%} & \multicolumn{3}{c}{99\%} \\
    \bottomrule
    \end{tabular}%
  \label{Perf_CCLM}%
\end{table*}%

\begin{table*}[htbp]
  \centering
  \caption{ Coding Performance of TSCPM on HPM-6.0 Testing Platform under AI, RA and LDB Configurations}
    \begin{tabular}{c|c|c|c|c|c|c|c|c|c|c}
    \toprule
    \multicolumn{2}{c|}{\multirow{2}[2]{*}{Sequences}} & \multicolumn{3}{c|}{AI} & \multicolumn{3}{c|}{RA} & \multicolumn{3}{c}{LDB} \\
\cmidrule{3-11}    \multicolumn{2}{c|}{} & \multicolumn{1}{c}{Y} & \multicolumn{1}{c}{U} & V & \multicolumn{1}{c}{Y} & \multicolumn{1}{c}{U} & V & \multicolumn{1}{c}{Y} & \multicolumn{1}{c}{U} & V \\
    \midrule
    \multirow{4}{*}{720p} & City   & \multicolumn{1}{c}{0.02\%} & \multicolumn{1}{c}{-0.46\%} & -0.24\% & \multicolumn{1}{c}{0.05\%} & \multicolumn{1}{c}{-2.45\%} & -0.84\% & \multicolumn{1}{c}{-0.20\%} & \multicolumn{1}{c}{-1.05\%} & 0.07\% \\
          & Crew   & \multicolumn{1}{c}{-0.27\%} & \multicolumn{1}{c}{-7.17\%} & -6.71\% & \multicolumn{1}{c}{-0.28\%} & \multicolumn{1}{c}{-6.15\%} & -5.82\% & \multicolumn{1}{c}{-0.07\%} & \multicolumn{1}{c}{-2.81\%} & -1.77\% \\
          & Vidyo1 & \multicolumn{1}{c}{-0.03\%} & \multicolumn{1}{c}{-5.12\%} & -3.34\% & \multicolumn{1}{c}{-0.03\%} & \multicolumn{1}{c}{-4.99\%} & -2.74\% & \multicolumn{1}{c}{0.19\%} & \multicolumn{1}{c}{-2.25\%} & -2.51\% \\
          & Vidyo3 & \multicolumn{1}{c}{0.02\%} & \multicolumn{1}{c}{-3.55\%} & -10.07\% & \multicolumn{1}{c}{-0.16\%} & \multicolumn{1}{c}{-4.90\%} & -13.46\% & \multicolumn{1}{c}{-0.31\%} & \multicolumn{1}{c}{-6.56\%} & -14.20\% \\
    \midrule
    \multirow{4}{*}{1080p} & BasketballDrive & \multicolumn{1}{c}{-0.46\%} & \multicolumn{1}{c}{-7.28\%} & -7.88\% & \multicolumn{1}{c}{-0.24\%} & \multicolumn{1}{c}{-7.38\%} & -6.86\% & \multicolumn{1}{c}{-0.11\%} & \multicolumn{1}{c}{-4.11\%} & -3.84\% \\
          & Cactus          & \multicolumn{1}{c}{-0.51\%} & \multicolumn{1}{c}{-9.59\%} & -8.67\% & \multicolumn{1}{c}{-0.28\%} & \multicolumn{1}{c}{-12.24\%} & -8.28\% & \multicolumn{1}{c}{0.02\%} & \multicolumn{1}{c}{-7.38\%} & -6.30\% \\
          & MarketPlace     & \multicolumn{1}{c}{-0.66\%} & \multicolumn{1}{c}{-13.85\%} & -15.14\% & \multicolumn{1}{c}{0.03\%} & \multicolumn{1}{c}{-15.82\%} & -14.41\% & \multicolumn{1}{c}{-0.07\%} & \multicolumn{1}{c}{-4.73\%} & -3.54\% \\
          & RitualDance     & \multicolumn{1}{c}{-0.39\%} & \multicolumn{1}{c}{-10.00\%} & -16.47\% & \multicolumn{1}{c}{-0.30\%} & \multicolumn{1}{c}{-5.50\%} & -13.08\% & \multicolumn{1}{c}{-0.06\%} & \multicolumn{1}{c}{-4.73\%} & -8.23\% \\
    \midrule
    \multirow{4}{*}{4K} & Tango2        & \multicolumn{1}{c}{-1.98\%} & \multicolumn{1}{c}{-26.86\%} & -27.70\% & \multicolumn{1}{c}{-0.76\%} & \multicolumn{1}{c}{-24.67\%} & -24.30\% & \multicolumn{1}{c}{-0.31\%} & \multicolumn{1}{c}{-14.45\%} & -12.36\% \\
          & Campfire      & \multicolumn{1}{c}{-9.28\%} & \multicolumn{1}{c}{-36.25\%} & -37.18\% & \multicolumn{1}{c}{-5.94\%} & \multicolumn{1}{c}{-20.37\%} & -26.56\% & \multicolumn{1}{c}{-6.17\%} & \multicolumn{1}{c}{-23.14\%} & -25.97\% \\
          & ParkRunning3  & \multicolumn{1}{c}{-1.00\%} & \multicolumn{1}{c}{-2.68\%} & -1.79\% & \multicolumn{1}{c}{-0.35\%} & \multicolumn{1}{c}{-1.54\%} & -0.82\% & \multicolumn{1}{c}{-0.08\%} & \multicolumn{1}{c}{-0.64\%} & -0.28\% \\
          & DaylightRoad2 & \multicolumn{1}{c}{-0.25\%} & \multicolumn{1}{c}{-11.61\%} & -3.66\% & \multicolumn{1}{c}{-0.14\%} & \multicolumn{1}{c}{-10.68\%} & -3.56\% & \multicolumn{1}{c}{0.02\%} & \multicolumn{1}{c}{-7.10\%} & -2.05\% \\
    \midrule
    \multicolumn{2}{c|}{720p} & \multicolumn{1}{c}{-0.06\%} & \multicolumn{1}{c}{-4.08\%} & -5.09\% & \multicolumn{1}{c}{-0.11\%} & \multicolumn{1}{c}{-4.62\%} & -5.71\% & \multicolumn{1}{c}{-0.10\%} & \multicolumn{1}{c}{-3.17\%} & -4.60\% \\
    \multicolumn{2}{c|}{1080p} & \multicolumn{1}{c}{-0.51\%} & \multicolumn{1}{c}{-10.18\%} & -12.04\% & \multicolumn{1}{c}{-0.20\%} & \multicolumn{1}{c}{-10.24\%} & -10.65\% & \multicolumn{1}{c}{-0.05\%} & \multicolumn{1}{c}{-5.24\%} & -5.48\% \\
    \multicolumn{2}{c|}{4K} & \multicolumn{1}{c}{-3.12\%} & \multicolumn{1}{c}{-19.35\%} & -17.58\% & \multicolumn{1}{c}{-1.80\%} & \multicolumn{1}{c}{-14.31\%} & -13.81\% & \multicolumn{1}{c}{-1.63\%} & \multicolumn{1}{c}{-11.33\%} & -10.17\% \\
    \midrule
    \multicolumn{2}{c|}{\textbf{Overall}} & \multicolumn{1}{c}{\textbf{-1.23\%}} & \multicolumn{1}{c}{\textbf{-11.20\%}} & \textbf{-11.57}\% & \multicolumn{1}{c}{\textbf{-0.70\%}} & \multicolumn{1}{c}{\textbf{-9.72\%}} & \textbf{-10.06\%} & \multicolumn{1}{c}{\textbf{-0.60\%}} & \multicolumn{1}{c}{\textbf{-6.58\%}} & \textbf{-6.75\%} \\
    \midrule
    \multicolumn{2}{c|}{Enc Time} & \multicolumn{3}{c|}{100\%} & \multicolumn{3}{c|}{100\%} & \multicolumn{3}{c}{99\%} \\
    \midrule
    \multicolumn{2}{c|}{Dec Time} & \multicolumn{3}{c|}{100\%} & \multicolumn{3}{c|}{100\%} & \multicolumn{3}{c}{99\%} \\
    \bottomrule
    \end{tabular}%
  \label{Perf_AVS}%
\end{table*}%

\begin{table*}[htbp]
  \centering
  \caption{Statistical Results of $\Delta E_{r}$ in Chroma CB under AI, RA and LDB Configurations}
    \begin{tabular}{cc|cccc|cccc|cccc}
    \toprule
    \multicolumn{2}{c|}{\multirow{2}[4]{*}{Sequences}} & \multicolumn{4}{c|}{AI}       & \multicolumn{4}{c|}{RA}       & \multicolumn{4}{c}{LDB} \\
\cmidrule{3-14}    \multicolumn{2}{c|}{} & 27    & 32    & 38    & 45    & 27    & 32    & 38    & 45    & 27    & 32    & 38    & 45 \\
    \midrule
    \multicolumn{1}{c|}{\multirow{4}[2]{*}{720p}} & City  & 1.8\% & 1.1\% & 1.3\% & 0.2\% & 1.0\% & 1.3\% & 0.9\% & 0.7\% & 0.3\% & 1.0\% & 0.7\% & 0.6\% \\
    \multicolumn{1}{c|}{} & Crew  & 6.7\% & 10.5\% & 15.1\% & 17.8\% & 1.1\% & 3.9\% & 2.5\% & 2.9\% & 1.6\% & 0.9\% & 1.8\% & 6.0\% \\
    \multicolumn{1}{c|}{} & Vidyo1 & 8.8\% & 11.5\% & 13.2\% & 13.1\% & 11.0\% & 14.2\% & 7.5\% & 6.8\% & 0.9\% & 10.1\% & 13.5\% & 4.9\% \\
    \multicolumn{1}{c|}{} & Vidyo3 & 25.4\% & 25.1\% & 20.3\% & 9.2\% & 14.6\% & 20.9\% & 20.3\% & 5.5\% & 13.1\% & 12.1\% & 8.0\% & 2.6\% \\
    \midrule
    \multicolumn{1}{c|}{\multirow{4}[2]{*}{1080p}} & BasketballDrive & 10.3\% & 13.5\% & 16.4\% & 17.5\% & 6.0\% & 6.7\% & 11.3\% & 12.1\% & 2.3\% & 5.2\% & 7.1\% & 5.9\% \\
    \multicolumn{1}{c|}{} & Cactus & 11.6\% & 14.1\% & 16.1\% & 17.7\% & 10.0\% & 7.4\% & 11.5\% & 18.0\% & 4.6\% & 6.4\% & 7.9\% & 10.1\% \\
    \multicolumn{1}{c|}{} & MarketPlace & 23.8\% & 27.0\% & 31.1\% & 30.5\% & 12.6\% & 17.7\% & 16.7\% & 22.8\% & 9.0\% & 10.3\% & 12.8\% & 12.3\% \\
    \multicolumn{1}{c|}{} & RitualDance & 19.5\% & 23.9\% & 26.5\% & 24.7\% & 12.4\% & 16.1\% & 18.9\% & 20.1\% & 11.8\% & 13.1\% & 15.5\% & 14.0\% \\
    \midrule
    \multicolumn{1}{c|}{\multirow{4}[2]{*}{4K}} & Tango2 & 36.2\% & 38.5\% & 39.0\% & 38.6\% & 29.6\% & 35.7\% & 33.7\% & 34.0\% & 20.8\% & 22.9\% & 24.6\% & 24.3\% \\
    \multicolumn{1}{c|}{} & Campfire & 52.6\% & 50.9\% & 43.0\% & 33.1\% & 33.9\% & 30.8\% & 25.1\% & 15.5\% & 30.7\% & 27.4\% & 20.5\% & 15.1\% \\
    \multicolumn{1}{c|}{} & ParkRunning3 & 2.4\% & 3.7\% & 5.7\% & 9.4\% & 1.0\% & 2.2\% & 4.1\% & 7.0\% & 0.4\% & 0.9\% & 1.7\% & 4.3\% \\
    \multicolumn{1}{c|}{} & DaylightRoad2 & 12.7\% & 15.3\% & 17.9\% & 17.6\% & 9.0\% & 13.7\% & 16.5\% & 17.6\% & 5.6\% & 9.8\% & 9.1\% & 13.0\% \\
    \midrule
    \multicolumn{2}{c|}{720P} & 10.7\% & 12.0\% & 12.5\% & 10.1\% & 6.9\% & 10.1\% & 7.8\% & 4.0\% & 4.0\% & 6.0\% & 6.0\% & 3.5\% \\
    \multicolumn{2}{c|}{1080P} & 16.3\% & 19.6\% & 22.5\% & 22.6\% & 10.3\% & 12.0\% & 14.6\% & 18.2\% & 6.9\% & 8.7\% & 10.8\% & 10.6\% \\
    \multicolumn{2}{c|}{4K } & 26.0\% & 27.1\% & 26.4\% & 24.7\% & 18.4\% & 20.6\% & 19.8\% & 18.5\% & 14.4\% & 15.2\% & 14.0\% & 14.2\% \\
    \midrule
    \multicolumn{2}{c|}{\textbf{Overall}} & \textbf{17.7\%} & \textbf{19.6\%} & \textbf{20.5\%} & \textbf{19.1\%} & \textbf{11.9\%} & \textbf{14.2\%} & \textbf{14.1\%} & \textbf{13.6\%} & \textbf{8.4\%} & \textbf{10.0\%} & \textbf{10.3\%} & \textbf{9.4\%} \\
    \bottomrule
    \end{tabular}%
  \label{resi}%
\end{table*}%

\begin{table*}[htbp]
  \centering
  \caption{The Operation Complexity of the Proposed Sub-sampled Cross-component Prediction}
    \begin{tabular}{c|c|c|c|c|c}
    \toprule
    Operations & Multiplication & Addition & Shift & Comparison & Down-sampling \\
    \midrule
    LSR~\cite{vvc2}   & 2$M$+4  & 7$M$+3  & 2     & -     & $M$ \\
    \midrule
    Max-Min~\cite{maxmin} & 1     & 3     & 1     & 2$M$    & $M$ \\
    \midrule
    Proposed & 1     & 7     & 5     & 4     & 4 \\
    \bottomrule
    \end{tabular}%
  \label{OpComplexity}%
\end{table*}%

\begin{table}[htbp]
  \centering
  \caption{Illustration of the Operation Complexity Regarding a $32\times 32$ Chroma CB}
    \begin{tabular}{c|cc}
    \toprule
      Operation    & Comparison &  Down-sampling \\
    \midrule
      $O$~\cite{maxmin}    & 128   & 64 \\
    \midrule
      $O^*$    & 4     & 4 \\
    \midrule
      $1-(O^* / O)$    & 97\%  & 94\% \\
    \bottomrule
    \end{tabular}%
  \label{32x32}%
\end{table}%

\section{Experimental Results}
The performance of the proposed sub-sampled cross-component prediction has been repeatedly validated in the standardization of VVC and AVS3 standards. Due to the excellent performance in balancing rate-distortion performance and computational complexity, both VVC and AVS3 adopt such strategy. In this section, the implementation details regarding the incorporation of the design philosophy of sub-sampled cross-component prediction in these standards, as well as the performance on the corresponding test models are presented.  

\subsection{Performance Evaluations on VVC}
 In VVC, besides the exact above and exact left reference samples, the samples located at above-right and below-left are also eligible to be employed. More specifically, supposing the chroma block size is $w\times h$, the available sample number in the above side $w'$ and left side $h'$ is given by,
\begin{equation}
    w'=
    \begin{cases}
        w, & \text{LM}\\
        w+\min(N_{ar}, h), &\text{LM-Above},
    \end{cases}
\end{equation}
\begin{equation}
    h'=
    \begin{cases}
        h, & \text{LM}\\
        h+\min(N_{bl}, h), &\text{LM-Left},
    \end{cases}
\end{equation}
where $N_{ar}$ and $N_{bl}$ denote the available sample number in the above right and below left.

The sub-sampled cross-component prediction scheme is first evaluated based on the VVC test model VTM-5.0~\cite{VTM5}. The JVET recommended sequences from class A1 to class F are involved in the simulation conforming to the common test conditions (CTC)~\cite{ctc5} under all intra (AI), random access (RA) and low delay B (LDB) configurations. The quantization parameters (QPs) are set as 22, 27, 32 and 37. BD-Rate~\cite{bjontegaard2001calculation} is employed for the coding performance evaluation, where the negative BD-Rate indicates performance improvement. Experimental results of the proposed scheme are illustrated in Table~\ref{Perf_VVC} in comparison with the Max-Min~\cite{maxmin} scheme where 0.06\%, 0.23\% and 0.29\% BD-Rate gains are achieved for Y, U and V components, respectively under AI configuration. Moreover, 0.05\%, 0.13\% and 0.33\% BD-Rate savings for Y, U and V components can be observed under RA configuration. The proposed scheme brings similar coding performance promotion under LDB configurations. Furthermore, with the cooperation of the sub-sampled scheme, CCLM achieves 1.76\%, 13.30\% and 13.83\% BD-Rate savings for Y, U and V components, respectively, under AI configuration. In case of RA configuration, the coding gain for luma component is 0.87\%, for chroma is 13.23\% (U) and 13.62\% (V). The coding performance under LDB configuration is 0.12\%, 4.28\% and 5.27\% for Y , U and V components, as illustrated in Table~\ref{Perf_CCLM}.

\subsection{Performance Evaluations on AVS3}
The core of TSCPM is grounded on the sub-sampled cross-component prediction, wherein the linear model parameters are derived according to the sub-sampled luma and chroma pairs. To economize the buffering resources, only the reference samples locating directly to the left and above sides of the current CB are accessible. The linear model is directly applied to the luma reconstructed CB that is with the original dimension, yielding a temporary chroma CB. Subsequently, down-sampling filter with coefficients $[1, 2, 1; 1, 2, 1]$ is applied to the temporary chroma CB, yielding the final chroma prediction block.

We verify the performance of TSCPM on the AVS3 test platform HPM-6.0~\cite{hpm6} following the CTCs~\cite{avs3ctc_hpm6}. The QPs are set as 22, 27, 32 and 37. The coding performance regarding individual test sequence is tabulated in Table~\ref{Perf_AVS}. We can notice that the TSCPM substantially improves the compression performance, especially in term of the rate-distortion performance of the chroma components. The TSCPM improves the prediction accuracy of chroma components, leading to the reduction of the residual energies in chroma CBs and overall coding bits, which in turn brings the performance improvement for the luma component. In particular, with AI configuration, the TSCPM achieves 1.23\% BD-Rate savings for luma component, as well as over 11\% BD-Rate savings for chroma components. In addition, under RA configuration there are 0.70\%, 9.72\% and 10.06\% BD-Rate gains on average for Y, U and V components, respectively . Moreover, the TSCPM is capable of achieving 0.60\%, 6.58\% and 6.75\% performance improvement for Y, U and V components under LDB configuration. It is interesting to see that the performance of the TSCPM is impressive for 4K sequences where over 3\% BD-Rate gains can be achieved on luma component, and nearly 20\% BD-Rate savings can be achieved on chroma components under AI configuration. The largest gain is from the sequence ``Campfire'', which is essentially a dark scene with flickering lights. Traditional intra prediction modes cannot well handle such a scenario whereas the TSCPM reveals its superiority by conducting the inter-channel prediction.

To further provide evidence regarding the capability in removing redundancy, we collect the residual energies for chroma components, which can be defined as,
\begin{align} 
    E_{r} = \sum_i (r_i)^2,
\end{align}
where $r_i$ represents the entropy-coded residuals. The variation of residual energies can be formulated as,
\begin{align}
    \Delta E_{r} = \frac{E_{r}^{anc} - E_{r}^{pro}}{E_{r}^{anc}} \times 100\%.
\end{align}
Herein, $E_{r}^{anc}$ and $E_{r}^{pro}$ denote the residual energies of the anchor and the proposed methods within chroma components, respectively. Positive value of $\Delta E_{r}$ indicates the reductions of residual energies. As illustrated in Table~\ref{resi}, it can be observed that on average 18\% to 21\% reductions of residual energies can be achieved with various QPs by TSCPM under AI configuration. The savings of residual energies imply that TSCPM efficiently removes the cross-component redundancy and improves the prediction accuracy for chroma components. In particular, the sequence ``Campfire'' with the largest gain achieves 33\% to 53\% savings of residual energies under AI configuration.

\subsection{Analyses of Operation Complexity}
Regarding the operation complexity of the linear model derivation, as illustrated in Table~\ref{OpComplexity}, the sub-sampled cross-component prediction significantly decreases the number of multiplication, addition and down-sampling operations when compared with the conventional LSR scheme. Supposing the number of the accessible reference sample pairs is $M$, the proposed sub-sampled scheme requires only one multiplication in deriving the parameter $\beta$, along with seven additions and five shift operations. Moreover, since at most four reference sample pairs are involved in the linear model derivation, with the proposed method, four comparisons are sufficient in discriminating the larger two and smaller two sample pairs. Furthermore, the number of down-sampling operations is four at most regardless of the block sizes. Therefore, for a typical $32 \times 32$ chroma CB, $M$ equals to 64 if only the exact above and left reference samples are explored. Moreover, we use $O$ to denote the explicit complexity of a specific kind of operation. The operation complexity $O$ can be concluded as 128 comparisons and 64 down-sampling operations when employing the Max-Min scheme to derive the linear model parameters. By contrast, with the proposed scheme, the associated operation complexity $O^*$ persists as four, such that 97\% comparisons and 94\% down-sampling operations can be eliminated in such a scenario, as illustrated in Table~\ref{32x32}.

\section{Conclusions}
We have proposed a sub-sampled cross-component prediction for the chroma intra coding. Aiming for alleviating the operational complexity for both encoder and decoder in deriving the chroma inference model, the proposed scheme is elaborated in a scientifically sound way from the perspective of the inter-pixel correlations and the inter-channel correlations mining. Instead of visiting all reference samples which leads to large quantities of multiplications, additions, comparisons or luma down-sampling, the proposed scheme tackles this problem by employing at most four luma and chroma sample pairs with fixed positions from the neighboring reconstructed sample set, which significantly mitigates the computational overheads. Extensive experimental results verify that the proposed linear model derivation scheme is corroborated to be robust and effective, leading to the adoption by the VVC and AVS3 standards.


%





\ifCLASSOPTIONcaptionsoff
  \newpage
\fi



\small
\bibliographystyle{IEEEtran}
\bibliography{refs}

\begin{thebibliography}{10}
\providecommand{\url}[1]{#1}
\csname url@samestyle\endcsname
\providecommand{\newblock}{\relax}
\providecommand{\bibinfo}[2]{#2}
\providecommand{\BIBentrySTDinterwordspacing}{\spaceskip=0pt\relax}
\providecommand{\BIBentryALTinterwordstretchfactor}{4}
\providecommand{\BIBentryALTinterwordspacing}{\spaceskip=\fontdimen2\font plus
\BIBentryALTinterwordstretchfactor\fontdimen3\font minus
  \fontdimen4\font\relax}
\providecommand{\BIBforeignlanguage}[2]{{%
\expandafter\ifx\csname l@#1\endcsname\relax
\typeout{** WARNING: IEEEtran.bst: No hyphenation pattern has been}%
\typeout{** loaded for the language `#1'. Using the pattern for}%
\typeout{** the default language instead.}%
\else
\language=\csname l@#1\endcsname
\fi
#2}}
\providecommand{\BIBdecl}{\relax}
\BIBdecl

\bibitem{bovik2010handbook}
A.~C. Bovik, \emph{Handbook of image and video processing}.\hskip 1em plus
  0.5em minus 0.4em\relax Academic press, 2010.

\bibitem{zhangkaiTIP}
K.~{Zhang}, J.~{Chen}, L.~{Zhang}, X.~{Li}, and M.~{Karczewicz}, ``Enhanced
  cross-component linear model for chroma intra-prediction in video coding,''
  \emph{IEEE Transactions on Image Processing}, vol.~27, no.~8, pp. 3983--3997,
  2018.

\bibitem{zhangxingyuTIP}
X.~{Zhang}, C.~{Gisquet}, E.~{François}, F.~{Zou}, and O.~C. {Au}, ``Chroma
  intra prediction based on inter-channel correlation for hevc,'' \emph{IEEE
  Transactions on Image Processing}, vol.~23, no.~1, pp. 274--286, 2014.

\bibitem{ZhangTao}
{T. Zhang}, {X. Fan}, {D. Zhao}, and {W. Gao}, ``Improving chroma intra
  prediction for {HEVC},'' in \emph{2016 IEEE International Conference on
  Multimedia Expo Workshops (ICMEW)}, 2016, pp. 1--6.

\bibitem{Liyue}
Y.~{Li}, L.~{Li}, Z.~{Li}, J.~{Yang}, N.~{Xu}, D.~{Liu}, and H.~{Li}, ``A
  hybrid neural network for chroma intra prediction,'' in \emph{2018 25th IEEE
  International Conference on Image Processing (ICIP)}, 2018, pp. 1797--1801.

\bibitem{M0263}
K.~Zhang, L.~Zhang, H.~Liu, J.~Xu, Y.~Wang, P.~Zhao, and D.~Hong, ``{CE3}:
  {CCLM} prediction with single-line neighbouring luma samples,''
  \emph{JVET-M0263}, Jan. 2019.

\bibitem{M0274}
M.~Wang, K.~Zhang, L.~Zhang, H.~Liu, J.~Xu, S.~Wang, J.~Li, S.~Wang, and
  W.~Gao, ``{CE3-related}: Modified linear model derivation for {CCLM} modes,''
  \emph{JVET-M0274}, Jan. 2019.

\bibitem{M0356}
A.~Filippov, X.~Ma, V.~Rufitskiy, H.~Yang, and J.~Chen, ``{CE3-related}:
  Simplified calculation for {CCLM} parameters derivation,'' \emph{JVET-M0356},
  Jan. 2019.

\bibitem{M0401}
M.~Ma, A.~Filippov, V.~Rufitskiy, H.~Yang, and J.~Chen, ``{CE3}:
  Classification-based mean value for {CCLM} coefficients derivation,''
  \emph{JVET-M0401}, Jan. 2019.

\bibitem{maxmin}
G.~Larche, J.~Taquet, C.~Gisquet, and P.~Onno, ``{CE3-5.1}: On cross-component
  linear model simplification,'' \emph{JVET-L0191}, Oct. 2018.

\bibitem{DCC_CCLM}
J.~{Li}, M.~{Wang}, L.~{Zhang}, K.~{Zhang}, S.~{Wang}, S.~{Wang}, S.~{Ma}, and
  W.~{Gao}, ``Sub-sampled cross-component prediction for chroma component
  coding,'' in \emph{2020 Data Compression Conference (DCC)}, 2020, pp.
  203--212.

\bibitem{vvc4}
B.~Bross, J.~Chen, and S.~Liu, ``Versatile video coding (draft 4),''
  \emph{JVET-M1001}, 2019.

\bibitem{AVS3_jiaqi}
J.~{Zhang}, C.~{Jia}, M.~{Lei}, S.~{Wang}, S.~{Ma}, and W.~{Gao}, ``Recent
  development of avs video coding standard: Avs3,'' in \emph{2019 Picture
  Coding Symposium (PCS)}, 2019, pp. 1--5.

\bibitem{MTT}
X.~Li, H.-C. Chuang, J.~Chen, M.~Karczewicz, L.~Zhang, X.~Zhao, and A.~Said,
  ``Multi-type-tree,'' \emph{Joint Video Exploration Team (JVET), doc.
  JVET-D0117}, 2016.

\bibitem{EQT_TIP}
M.~{Wang}, J.~{Li}, L.~{Zhang}, K.~{Zhang}, H.~{Liu}, S.~{Wang}, S.~{Kwong},
  and S.~{Ma}, ``Extended coding unit partitioning for future video coding,''
  \emph{IEEE Transactions on Image Processing}, vol.~29, pp. 2931--2946, 2020.

\bibitem{AMC}
K.~{Zhang}, Y.~{Chen}, L.~{Zhang}, W.~{Chien}, and M.~{Karczewicz}, ``An
  improved framework of affine motion compensation in video coding,''
  \emph{IEEE Transactions on Image Processing}, vol.~28, no.~3, pp. 1456--1469,
  2019.

\bibitem{ISP_VVC}
S.~{De-Luxán-Hernández}, V.~{George}, J.~{Ma}, T.~{Nguyen}, H.~{Schwarz},
  D.~{Marpe}, and T.~{Wiegand}, ``An intra subpartition coding mode for
  {VVC},'' in \emph{2019 IEEE International Conference on Image Processing
  (ICIP)}, 2019, pp. 1203--1207.

\bibitem{WideAngular}
L.~{Zhao}, X.~{Zhao}, S.~{Liu}, X.~{Li}, J.~{Lainema}, G.~{Rath}, F.~{Urban},
  and F.~{Racapé}, ``Wide angular intra prediction for versatile video
  coding,'' in \emph{2019 Data Compression Conference (DCC)}, 2019, pp. 53--62.

\bibitem{AMVR}
H.~{Liu}, L.~{Zhang}, K.~{Zhang}, J.~{Xu}, Y.~{Wang}, J.~{Luo}, and Y.~{He},
  ``Adaptive motion vector resolution for affine-inter mode coding,'' in
  \emph{2019 Picture Coding Symposium (PCS)}, 2019, pp. 1--4.

\bibitem{HMVP_ICME}
J.~{Li}, M.~{Wang}, L.~{Zhang}, K.~{Zhang}, H.~{Liu}, S.~{Wang}, S.~{Ma}, and
  W.~{Gao}, ``History-based motion vector prediction for future video coding,''
  in \emph{2019 IEEE International Conference on Multimedia and Expo (ICME)},
  2019, pp. 67--72.

\bibitem{HMVP_DCC}
L.~{Zhang}, K.~{Zhang}, H.~{Liu}, H.~C. {Chuang}, Y.~{Wang}, J.~{Xu},
  P.~{Zhao}, and D.~{Hong}, ``History-based motion vector prediction in
  versatile video coding,'' in \emph{2019 Data Compression Conference (DCC)},
  2019, pp. 43--52.

\bibitem{FIMC_TIP}
J.~{Li}, M.~{Wang}, L.~{Zhang}, K.~{Zhang}, H.~{Liu}, S.~{Wang}, S.~{Ma}, and
  W.~{Gao}, ``Unified intra mode coding based on short and long range
  correlations,'' \emph{IEEE Transactions on Image Processing}, pp. 1--1, 2020.

\bibitem{Transform_Zhaoxin}
X.~{Zhao}, J.~{Chen}, M.~{Karczewicz}, A.~{Said}, and V.~{Seregin}, ``Joint
  separable and non-separable transforms for next-generation video coding,''
  \emph{IEEE Transactions on Image Processing}, vol.~27, no.~5, pp. 2514--2525,
  2018.

\bibitem{CCLM2010}
J.~Kim, S.-W. Park, J.-Y. Park, and B.-M. Jeon, ``Intra chroma prediction using
  inter channel correlation,'' \emph{JCTVC-B021}, Jul. 2010.

\bibitem{CCLM2010_Oct}
J.~Chen and V.~Seregin, ``Chroma intra prediction by scaled luma samples using
  integer operations,'' \emph{JCTVC-C206}, Oct. 2010.

\bibitem{CCP_HEVC}
W.~{Kim}, W.~{Pu}, A.~{Khairat}, M.~{Siekmann}, J.~{Sole}, J.~{Chen},
  M.~{Karczewicz}, T.~{Nguyen}, and D.~{Marpe}, ``Cross-component prediction in
  {HEVC},'' \emph{IEEE Transactions on Circuits and Systems for Video
  Technology}, pp. 1--1, 2015.

\bibitem{rudat2019inter}
C.~Rudat, C.~R. Helmrich, J.~Lainema, T.~Nguyen, H.~Schwarz, D.~Marpe, and
  T.~Wiegand, ``Inter-component transform for color video coding,'' in
  \emph{2019 Picture Coding Symposium (PCS)}.\hskip 1em plus 0.5em minus
  0.4em\relax IEEE, 2019, pp. 1--5.

\bibitem{HEVCREOverview}
D.~Flynn, D.~Marpe, M.~Naccari, T.~Nguyen, C.~Rosewarne, K.~Sharman, J.~Sole,
  and J.~Xu, ``Overview of the range extensions for the hevc standard: Tools,
  profiles, and performance,'' \emph{IEEE Transactions on Circuits and Systems
  for Video Technology}, vol.~26, no.~1, pp. 4--19, 2015.

\bibitem{jem4}
``{JEM4.0},'' \url{https://jvet.hhi.
  fraunhofer.de/svn/svn_HMJEMSoftware/tags/HM-16.6-JEM-4.0}, Oct. 2016.

\bibitem{vtm3}
``{VVC} software {VTM}-3.0,''
  \url{https://vcgit.hhi.fraunhofer.de/jvet/VVCSoftware_VTM/tags/VTM-3.0/}.

\bibitem{vvc2}
B.~Bross, J.~Chen, and S.~Liu, ``Versatile video coding (draft 2),''
  \emph{JVET-K1001}, 2018.

\bibitem{mauersberger1979generalised}
W.~Mauersberger, ``Generalised correlation model for designing 2-dimensional
  image coders,'' \emph{Electronics letters}, vol.~15, no.~20, pp. 664--665,
  1979.

\bibitem{AngularIntra_TIP2014}
J.~{Prades Nebot}, ``On the efficiency of angular intraprediction,'' \emph{IEEE
  Transactions on Image Processing}, vol.~23, no.~12, pp. 5733--5742, 2014.

\bibitem{VTM5}
``{VVC} software {VTM}-5.0,''
  \url{https://vcgit.hhi.fraunhofer.de/jvet/VVCSoftware_VTM/tags/VTM-5.0/}.

\bibitem{ctc5}
F.~Bossen, J.~Boyce, K.~Suehring, X.~Li, and V.~Seregin, ``{JVET} common test
  conditions and software reference configurations for {SDR} video,''
  \emph{Joint Video Exploration Team (JVET), doc. JVET-N1010}, Mar. 2019.

\bibitem{bjontegaard2001calculation}
G.~Bj{\o}ntegaard, ``Calculation of average {PSNR} differences between
  {RD}-curves,'' \emph{ITU-T SG 16 Q.6 VCEG-M33}, 2001.

\bibitem{hpm6}
``{AVS3} software repository,'' \url{https://gitlab.com/AVS3_Software/hpm.git}.

\bibitem{avs3ctc_hpm6}
J.~Chen, ``{AVS3-P2} common test conditions v9.0,'' \emph{AVS-N2762}, Jan.
  2020.

\end{thebibliography}
\end{document}